\pdfoutput=1
\newcommand*{\ATLASLATEXPATH}{latex/}
\documentclass[cernpreprint,texlive=2016,txfonts,UKenglish]{latex/atlasdoc}
\usepackage{\ATLASLATEXPATH atlaspackage}
\usepackage{\ATLASLATEXPATH atlasbiblatex}
\usepackage{\ATLASLATEXPATH atlascontribute}
\usepackage{\ATLASLATEXPATH atlasphysics}
\graphicspath{{logos/}{figures/}}

\usepackage{paper-defs}

\AtlasTitle{Study of \wwy and \wzy production in \pp collisions at $\sqrt{s} = \SI{8}{\TeV}$ and search for anomalous quartic 
gauge couplings with the ATLAS experiment}
\author{The ATLAS Collaboration}

\AtlasRefCode{ATLAS-STDM-2016-05}

\PreprintIdNumber{CERN-EP-2017-096}

\AtlasJournal{European Physical Journal C}
\AtlasJournalRef{{\EPJC} 77, 646 (2017)}
\AtlasDOI{10.1140/epjc/s10052-017-5180-3}

\AtlasAbstract{%
This paper presents a study of \wwy and \wzy triboson production using events from proton--proton collisions at a centre-of-mass 
energy of \mbox{$\sqrt{s} = \mbox{8\,\TeV}$} recorded with the ATLAS detector at the LHC and corresponding to an integrated 
luminosity of \SI{20.2}{\per\fb}.
The \wwy production cross-section is determined using a final state containing an electron, a muon, a photon, and neutrinos 
(\emy).
Upper limits on the production cross-section of the \emy final state and the \wwy and \wzy final states containing an electron or 
a muon, two jets, a photon, and a neutrino (\ejjy or \mjjy) are also derived. 
The results are compared to the cross-sections predicted by the Standard Model at next-to-leading order in the strong-coupling 
constant.
In addition, upper limits on the production cross-sections are derived in a fiducial region optimised for a search for new physics
beyond the Standard Model.
The results are interpreted in the context of anomalous quartic gauge couplings using an effective field theory.
Confidence intervals at 95\% confidence level are derived for the 14 coupling coefficients to which \wwy and \wzy 
production are 
sensitive.}

\hypersetup{pdftitle={ATLAS document},pdfauthor={The ATLAS Collaboration}}
\begin{document}

\maketitle


\section{Introduction}
\label{sec:intro}
Measuring triboson final states at the Large Hadron Collider (LHC)~\cite{Evans:2008zzb} provides a test of the non-Abelian 
structure of the electroweak sector of the Standard Model (SM) of particle physics that predicts quartic gauge couplings. 
Deviations from the SM can be parametrised in the framework of anomalous quartic gauge couplings (aQGCs). 
This paper describes a measurement of \wvy production by analysing events containing a \Wb boson, a vector boson~(\Vb),
being either another \Wb~boson or a \Zb~boson, and a photon, using proton--proton collisions at a centre-of-mass
energy of \mbox{$\sqrt{s} = 8\,\TeV$} corresponding to an integrated luminosity of \SI{20.2}{\per\fb} recorded by the ATLAS
detector~\cite{PERF-2007-01}.

At LEP, \wwy production was studied at centre-of-mass energies ranging from \mbox{183\,\GeV} to \mbox{207\,\GeV} in a variety of 
photon plus leptonic or hadronic final states~\cite{LEPelectroweak}. 
The analysis presented here has a higher energy reach than the results obtained at LEP.
The production of \wvy events was studied by the CMS Collaboration in Ref.~\cite{CMS-SMP-13-009} in final states containing
electrons or muons and jets, and using a data set with a similar luminosity and the same centre-of-mass energy as employed
here.
Other analyses with three bosons in the final state and also sensitive to quartic gauge couplings have been performed by the ATLAS 
and the CMS collaborations~\cite{STDM-2013-05, STDM-2014-01, STDM-2015-07, CMS-SMP-15-008}. 
Furthermore, exclusion limits on new physics beyond the SM described by aQGCs have also been set at the LHC using diboson final
states including photons~\cite{CMS-SMP-14-018, CMS-SMP-14-011,STDM-2015-21} and in diboson final states including
massive gauge bosons only~\cite{CMS-SMP-13-015, CMS-FSQ-13-008, STDM-2014-05, STDM-2013-06, STDM-2015-09,
STDM-2015-10}. 

\begin{figure}[htbp]
 \centering
 \subfloat[]{\includegraphics[width=.33\textwidth]{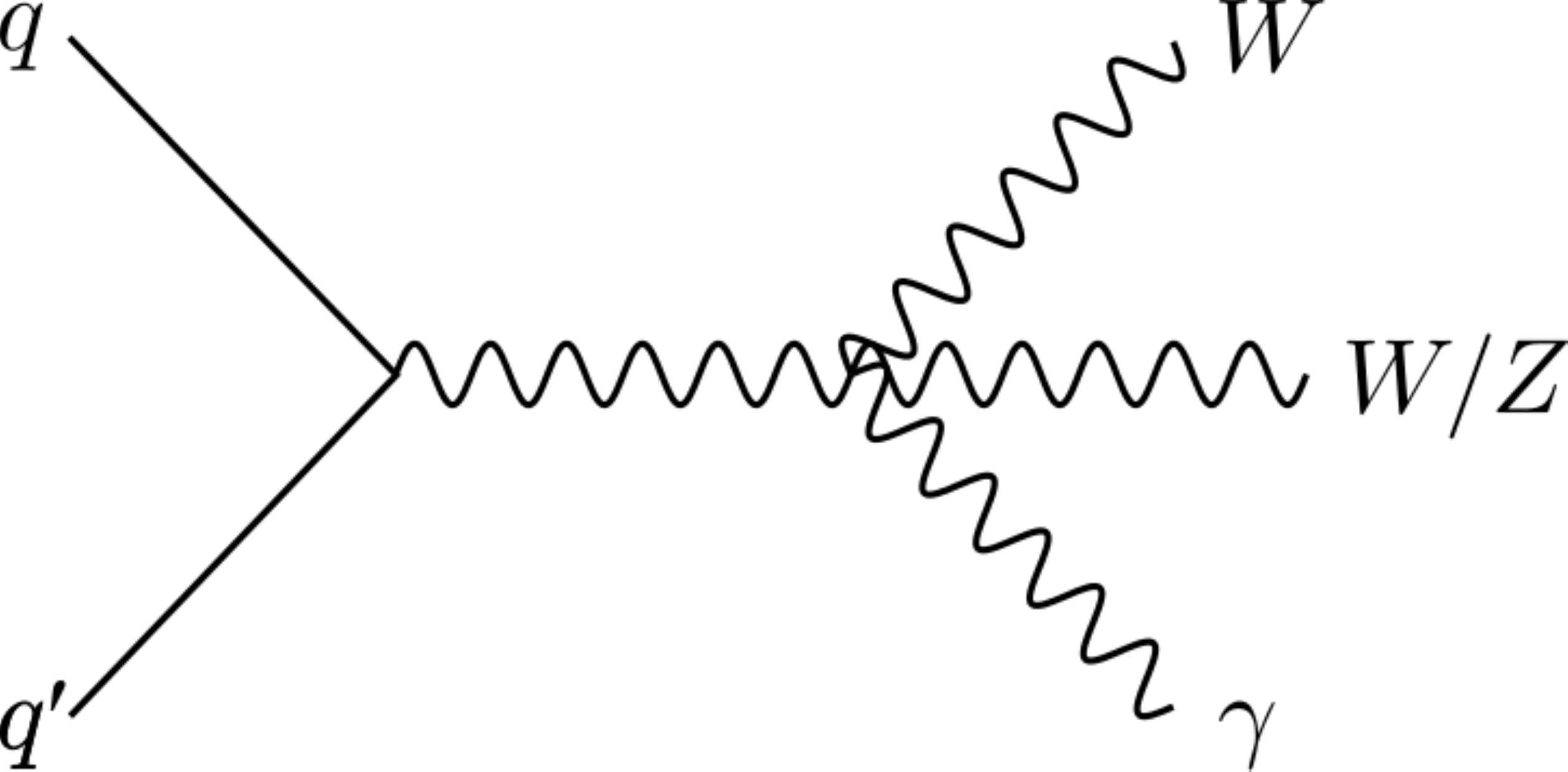}\label{fig:quartFeynm}} 
 \subfloat[]{\includegraphics[width=.33\textwidth]{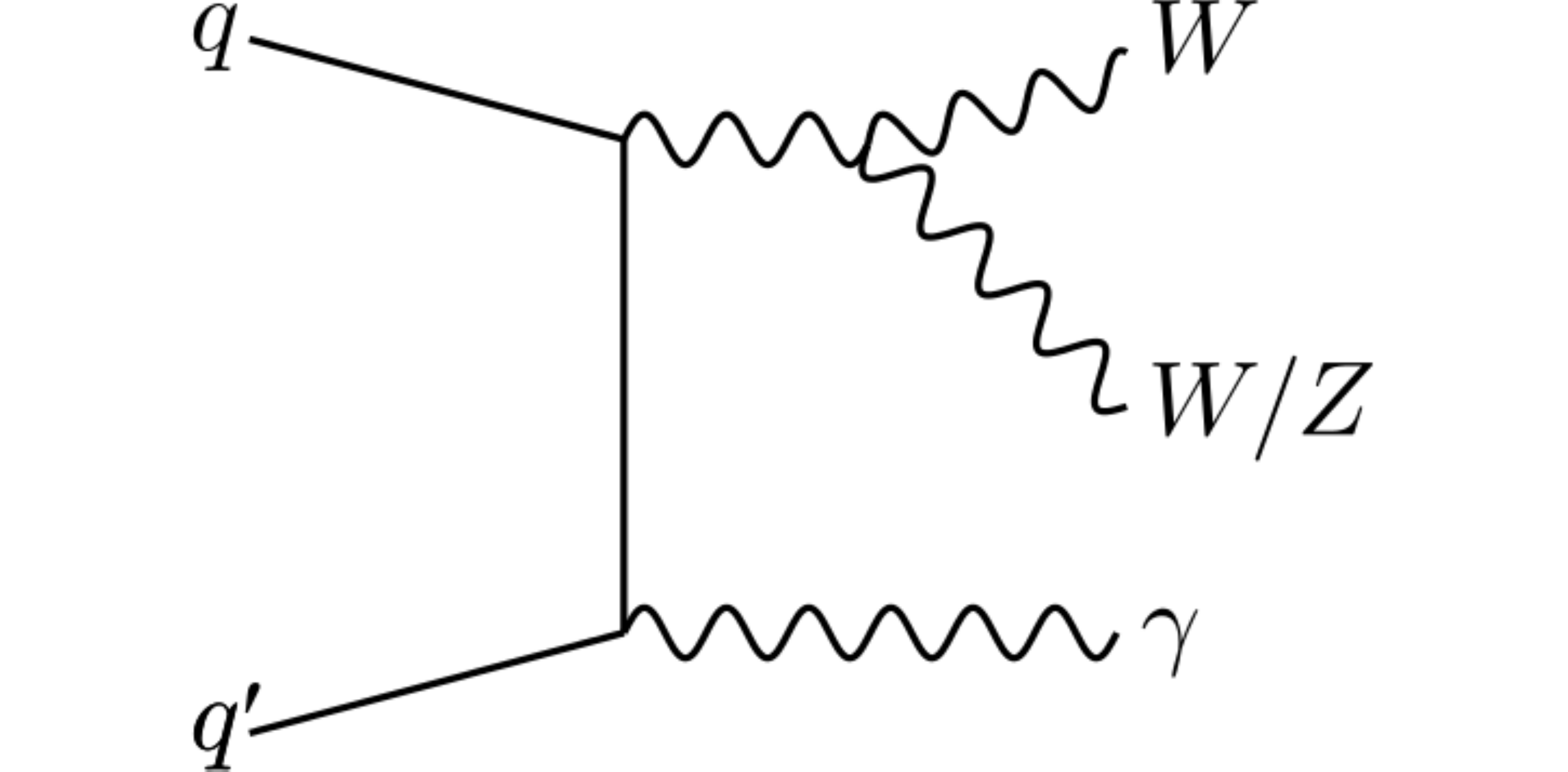}\label{fig:triplFeynm}}
 \subfloat[]{\includegraphics[width=.33\textwidth]{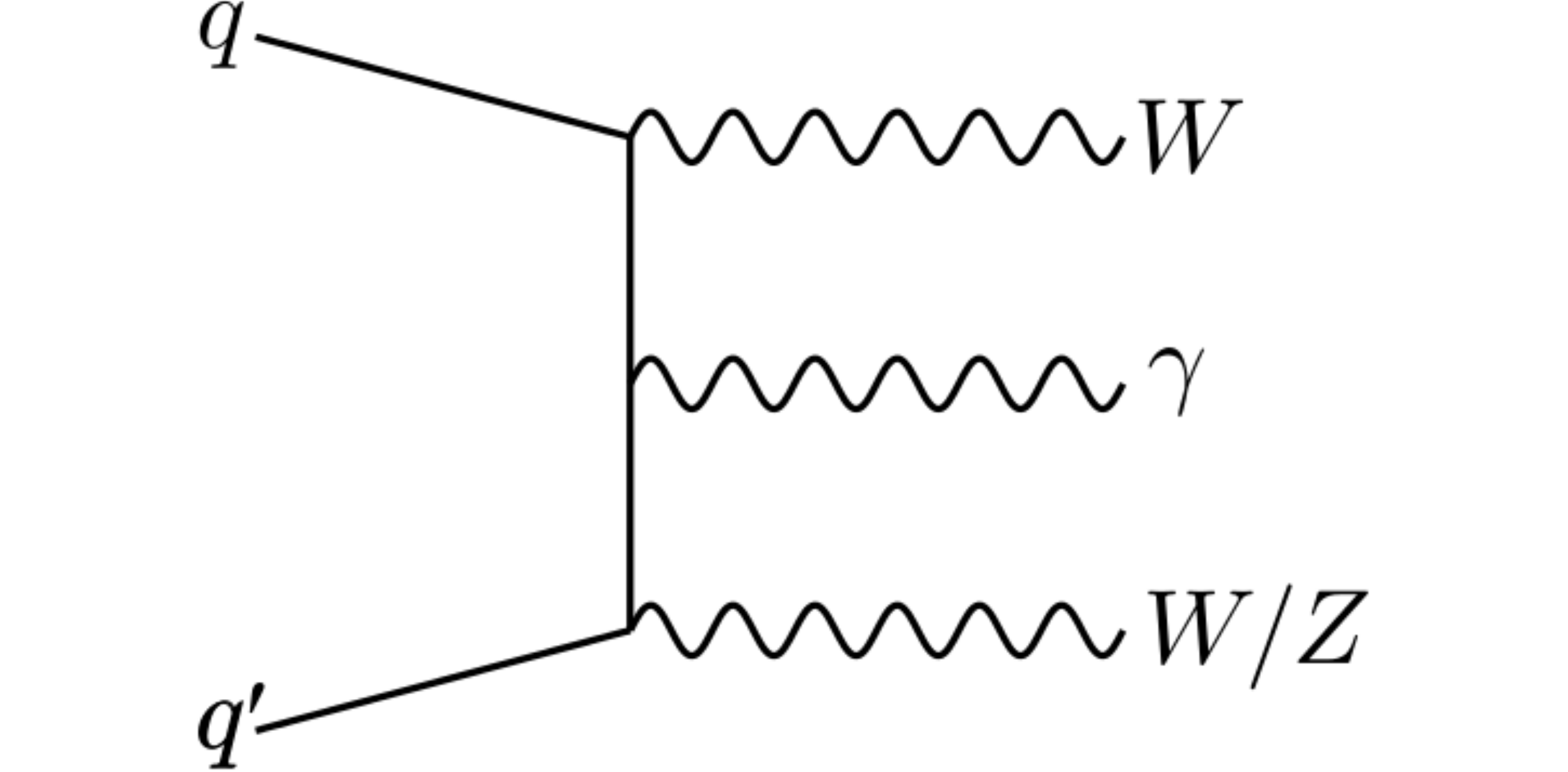}\label{fig:radiFeynm}}
 \caption{Examples of Feynman diagrams of \wvy production at the LHC. In \protect\subref{fig:quartFeynm} the quartic vertex is
shown, while \protect\subref{fig:triplFeynm} and \protect\subref{fig:radiFeynm} depict the production from radiative processes.}
  \label{fig:feynmans}
\end{figure}

In proton--proton collisions, \wvy events are produced through the $WWZ\gamma$ and $WW\gamma\gamma$ quartic couplings as
depicted in Figure~\ref{fig:quartFeynm} or through radiation of one or more bosons as exemplified in Figures~\ref{fig:triplFeynm}
and~\ref{fig:radiFeynm}. 
The fully leptonic final state~(\emy) of \wwy production containing an electron ($e$), a muon ($\mu$), their corresponding
neutrinos ($\nu$), and a photon is studied as it has a clean experimental signature. 
The same-flavour final states, $e\nu e\nu \gamma$ and $\mu\nu\mu\nu\gamma$, are not studied as they have large backgrounds.
Semileptonic final states~(\ljjy) containing one light lepton~($\ell = e \text{ or } \mu$), a neutrino, two
jets~($j$), and a photon are also studied. The analysis of the latter profits from the larger hadronic branching ratio of \Wb- and
\Zb-boson decays and is performed separately in the electron (\ejjy) and the muon (\mjjy) channels.
The production of \wvy events whose decays include \mbox{$\tau$ leptons} is not considered as signal.

Two fiducial regions are defined for all final states: 
one is optimised for the observation of the process while the other is optimised for a search for new physics beyond the SM. 
The results obtained in the latter region are interpreted in the context of aQGCs that describe modified triboson production
using an effective field theory~\cite{PhysRevD.74.073005}. 

This paper is structured as follows. 
The ATLAS detector and the data employed in this analysis are described in Section~\ref{sec:detector}. 
Section~\ref{sec:analysis} details the Monte Carlo simulations used. 
The reconstruction of the detector information is outlined in Section~\ref{sec:reconstruction}. 
The analysis of the fully leptonic final state is described in Section~\ref{sec:fully} followed by the description of the 
semileptonic analysis in Section~\ref{sec:semi}. 
In Section~\ref{sec:crosssec} the fiducial region of the cross-section measurement is defined and the determination of the 
production cross-section in the \emy final state is described. 
The derivation of upper limits on the \wvy production cross-section is also presented. 
Section~\ref{sec:aqgc} discusses the cross-section exclusion limits in the fiducial region optimised for new physics 
beyond the SM and the interpretation of the results in the framework of aQGCs. 
A summary of the results is given in Section~\ref{sec:conclusion}.

\section{ATLAS detector and data sample}
\label{sec:detector}
\newcommand{\AtlasCoordFootnote}{%
ATLAS uses a right-handed coordinate system with its origin at the nominal interaction point (IP) in the centre of the detector
and the $z$-axis along the beam line. The $x$-axis points from the IP to the centre of the LHC ring, and the $y$-axis points
upwards. Cylindrical coordinates $(r,\phi)$ are used in the transverse plane, $\phi$ being the azimuthal angle around the
$z$-axis. The rapidity ($y$) is defined as $y = \frac{1}{2}\ln\left[\frac{E+p_{z}}{E-p_{z}}\right]$, where $p_{z}$ is the
$z$-component of the momentum and $E$ is the energy of the object. The pseudorapidity ($\eta$) is defined in terms of the polar
angle $\theta$ as $\eta = -\ln \tan(\theta/2)$. Angular distance is measured in units of $\Delta R \equiv \sqrt{(\Delta\eta)^{2} +
(\Delta\phi)^{2}}$.}

The ATLAS experiment~\cite{PERF-2007-01} at the LHC is a multipurpose particle detector with a forward-backward symmetric
cylindrical geometry and a near $4\pi$ coverage in solid angle.\footnote{\AtlasCoordFootnote} 
It consists of an inner tracking detector surrounded by a thin superconducting solenoid providing a \SI{2}{\tesla} axial magnetic
field, electromagnetic and hadronic calorimeters, and a muon spectrometer. The inner tracking detector covers the pseudorapidity
range $|\eta| < 2.5$ and consists of silicon pixel, silicon microstrip, and transition radiation tracking detectors.
Lead/liquid-argon (LAr) sampling calorimeters provide electromagnetic energy measurements with high granularity in the
$\eta$--$\phi$ plane and a threefold segmentation in the radial direction. The first of the three layers of the LAr calorimeter 
has the smallest $\eta$-segmentation to discriminate between single photon showers and two overlapping showers coming from the 
decays of neutral hadrons. 
A hadronic (steel/scintillator-tile) calorimeter covers the central pseudorapidity range. The endcap and forward regions are
instrumented with LAr calorimeters for the energy measurement of electromagnetic and hadronic showers up to $|\eta| =
4.9$. The muon spectrometer encompasses the calorimeters and includes a system of precision tracking chambers as well as fast
detectors for triggering. 
It comprises three large air-core toroidal superconducting magnets with eight coils each. 
The field integral of the toroids ranges between \num{2.0} and \SI{6.0}{Tm} across most of the detector.
A three-level trigger system is used to select events for read-out and storage. 
The first-level trigger is implemented in hardware and uses a subset of the detector information to reduce the accepted rate to
\SI{75}{\kilo\hertz}. 
This is followed by two software-based trigger levels that together reduce the accepted event rate to \SI{400}{\hertz} on average.

This analysis uses data recorded at a centre-of-mass energy of $\mbox{8\,\TeV}$, corresponding to an integrated luminosity of
$20.2\,\pm\,0.4\,\text{fb}^{-1}$~\cite{DAPR-2013-01} after applying basic data quality criteria to ensure the full
functionality of all detector subcomponents. 
Only events that have at least three reconstructed tracks~\cite{ATLAS-CONF-2012-042} with \mbox{$\pt > 500\,\MeV$} associated with the primary vertex are
considered for analysis. 
The primary vertex is defined as the vertex whose associated tracks have the largest sum of squared transverse momenta.
Furthermore, events are discarded if they contain jets that are likely to be mismeasured.

Dedicated triggers are used for each final state. 
The events of the fully leptonic analysis are triggered by requiring three particles in the event: a muon with a transverse 
momentum (\pt) of at least \mbox{18\,\GeV} and two clusters of energy deposits in the electromagnetic calorimeter with a 
transverse energy ($\et$) of at least \mbox{10\,\GeV}. 
The efficiency of this trigger for the selection of the signal described in Section~\ref{sec:fully}
corresponds to \mbox{$0.82 \pm 0.01$(stat.)}.
For the semileptonic final states, a combination of single-lepton triggers~\cite{PERF-2011-02} is used to maintain a high efficiency over a wide
range of lepton transverse momenta. 
The \ejjy final state is triggered by either requiring an isolated electron with \mbox{$\pt > \mbox{24\,\GeV}$} or an electron
with \mbox{$\pt > \mbox{60\,\GeV}$} and no requirement on isolation.
The lepton isolation is based on the sum of the transverse momenta of additional tracks in a cone of size 
\mbox{$\Delta R = 0.2$} around the lepton's track.
This trigger combination provides an efficiency of \mbox{$0.964 \pm 0.004$(stat.)} for the signal selection described in 
Section~\ref{sec:semi}.
Similarly, the \mjjy final state is triggered by either requiring an isolated muon with \mbox{$\pt > \mbox{24\,\GeV}$} or a muon 
with \mbox{$\pt > \mbox{36\,\GeV}$} and no requirement on isolation. The efficiency of this trigger combination for the
signal corresponds to \mbox{$0.772 \pm 0.007$(stat.)}.

\section{Monte Carlo simulations}
\label{sec:analysis}
The expected signal and background events were simulated with Monte Carlo (MC) event generators. The simulations were used to
optimise the selection criteria, to compute efficiencies, and to estimate the contributions of specific background processes.
For the simulation of the MC samples, the ATLAS simulation infrastructure~\cite{SOFT-2010-01}, which uses the GEANT4
toolkit~\cite{Agostinelli:2002hh} for the detector simulation, was employed. 
All simulations described in this section were computed at leading order (LO) in the perturbative expansion of the
strong-coupling constant (\alphas) unless otherwise stated.

The \wvy signal process was simulated with the MC event generator SHERPA~2.1.1~\cite{Gleisberg:2008ta, Hoeche:2009rj, 
Gleisberg:2008fv, Schumann:2007mg} with up to one additional parton in the matrix element, using the default tunes. The CT10NLO~\cite{ct10} set of parton 
distribution functions (PDF) was used.
These signal predictions were normalised using the cross-sections of the fiducial regions introduced in 
Section~\ref{sec:crosssec},
computed at next-to-leading order (NLO) in \alphas using the VBFNLO~2.7.1~\cite{vbfnlo1, vbfnlo2,vbfnlo3, wwyQCD} program and the
CT14NLO~\cite{Dulat:2016} PDF set. 
The renormalisation and factorisation scales were set to the invariant mass of the triboson system. 
The \wvy processes that contain \mbox{$\tau$ leptons} in their decay are considered as background in this analysis and were
simulated like the signal as just described.
For cross-checks and for the estimation of systematic uncertainties associated with the event generation, the \wvy signal process
was also simulated using the MadGraph~5.2.2.2~\cite{Alwall:2014hca} event generator with dynamical renormalisation and 
factorisation scales.
It was interfaced to the PYTHIA~6.427~\cite{Sjostrand:2006za} program for the hadronisation and underlying event simulation with the Perugia 2012~\cite{Skands:2010ak} tune and
used the CTEQ6L1~\cite{Pumplin:2002vw} PDF set.
In addition, five reference samples modelling anomalous quartic gauge couplings were simulated for each studied final state, using
the MadGraph event generator as described above and normalised using the corresponding cross-section predictions obtained at NLO 
with the VBFNLO program.

Backgrounds from $WZ$, $ZZ$, and \zy diboson production were simulated with up to three additional partons in the final state
using the SHERPA event generator (versions 1.4.1, 1.4.5, and 1.4.1 with the default tunes respectively) with the CT10NLO PDF set.
Top quark pair production in association with a photon ($t\bar{t}\gamma$) was generated with the MadGraph~5.2.1.0 event generator 
using the CTEQ6L1 PDF set and interfaced to \mbox{PYTHIA 8.183}~\cite{Sjostrand:2007gs} for the simulation of the hadronisation 
and the underlying event using the AUET2B~\cite{ATL-PHYS-PUB-2011-009} tune. 
The cross-section was normalised using the computations of Ref.~\cite{PhysRevD.83.074013} which were performed at NLO in \alphas.
The simultaneous production of top and antitop quarks ($t\bar{t}$) and the production of \Wb bosons in association with top
quarks ($Wt$) were generated at NLO in \alphas with the POWHEG-BOX~\cite{Nason:2004rx, Frixione:2007vw,Re:2010bp} program using 
the CT10f4 PDF set and being interfaced to PYTHIA~6.426 with the Perugia 2011C~\cite{Skands:2010ak} tune and using the CTEQ6L1 PDF set.
The background from \Zb bosons produced in association with jets (\mbox{\Zb + jets}) and from \Wb-boson production in association
with a photon (\mbox{\wg+ jets}) were generated with the ALPGEN~\cite{Mangano:2002ea} program interfaced to the 
HERWIG~6.520.2~\cite{Corcella:2000bw} event generator for parton showering and hadronisation and to the JIMMY~\cite{Butterworth:1996zw} event generator to simulate the underlying event. 
The AUET2~\cite{ATL-PHYS-PUB-2011-008} tune and the CTEQ6L1 PDF set were employed.
All simulations that used the PYTHIA event generator employed the TAUOLA~\cite{Jadach:1993hs} program to compute the 
\mbox{$\tau$ lepton} decays. 
In samples that do not contain a prompt photon in the final state, the PHOTOS~\cite{Golonka:2005pn} program was employed to 
simulate photon radiation from final-state charged particles.

Contributions from additional proton--proton collisions accompanying the hard-scatter interaction, termed pile-up, were
simulated using the PYTHIA~8.160 event generator. 
The resulting distribution of the mean number of interactions per bunch crossing was corrected to reproduce the distribution
measured in data. 
The level of agreement between simulated and recorded data was further improved by correcting the simulated vertex distribution, 
object trigger and identification efficiencies, resolution and calibration to agree with the measured values~\cite{PERF-2016-01, 
PERF-2013-04, PERF-2014-05}.

\section{Event reconstruction}
\label{sec:reconstruction}
The selection of the \wvy signal events is based on objects that are reconstructed using the same algorithms for simulated 
and recorded events. The reconstruction of electron and photon candidates employs energy clusters~\cite{ATL-LARG-PUB-2008-002}
of the calorimeters and their matching to tracks from the inner detector~\cite{PERF-2016-01, ATL-PHYS-PUB-2011-007}. 
The measured energies of the electrons and photons are corrected as described in Ref.~\cite{PERF-2013-05}.
Electron or photon candidates reconstructed within \mbox{$1.37 < |\eta| < 1.52$} are discarded as this corresponds to a 
transition region between different calorimeter components which has poor energy resolution and identification efficiencies for 
these objects.

Photon candidates are reconstructed within $|\eta| < 2.37$ and their transverse energy has to exceed \mbox{15\,\GeV}.
They are required to fulfil the \emph{tight} identification criteria described in Ref.~\cite{PERF-2013-04}. 
An isolation requirement is applied to reject hadronic backgrounds: the additional transverse energy deposited in the calorimeter 
in a cone of size \mbox{$\Delta R = 0.4$} around the photon candidate, called \iso, must be less than \mbox{4\,\GeV} after the 
median energy density of the event scaled to the cone size is subtracted in order to reduce the effect from 
pile-up~\cite{STDM-2010-08}.

Electron candidates are reconstructed within $|\eta| < 2.47$ and their transverse momentum has to exceed \mbox{7\,\GeV}.
They are required to fulfil the \emph{tight} identification criteria described in Ref.~\cite{PERF-2016-01}.
In the fully leptonic analysis the same isolation requirement used for photons is applied to electrons as
this facilitates the background estimation with the two-dimensional sideband method (see Section~\ref{sec:fully}). 
The semileptonic analysis imposes a different isolation requirement, as it relies on other background estimation methods (see
Section~\ref{sec:semi}).
For this analysis, the additional transverse energy deposited in the calorimeter in a cone of size \mbox{$\Delta R = 0.3$} around 
the electron is required to be less than 14\% of the transverse energy of the electron after the pile-up energy is subtracted as 
for the photons.
Furthermore, a track-based isolation requirement is imposed: the sum of the transverse momenta of the additional tracks in the
aforementioned cone is required to be less than 7\% of the transverse energy of the electron itself. 
In addition, the semileptonic analysis requires the electron track to be consistent with coming from the primary vertex. 

Muon candidates are reconstructed within $|\eta| < 2.4$ by combining tracks in the inner detector with tracks in the muon 
spectrometer. 
A statistical combination of the track parameters or a global refit of the tracks, described as \emph{Chain 3} in 
Ref.~\cite{PERF-2014-05}, is used. 
Muon candidates are required to have a transverse momentum larger than \mbox{7\,\GeV} and to originate from the primary
vertex.
A track-based isolation requirement is imposed: 
the sum of the transverse momenta of the additional tracks in a cone of size \mbox{$\Delta R = 0.2$} around the muon candidate is
required to be less than 10\% of the transverse momentum of the muon candidate itself.

Jet candidates are reconstructed within $|y| < 4.4$ from topological energy clusters~\cite{PERF-2014-07} using the \antikt 
algorithm~\cite{Cacciari:2008gp} with a radius parameter of \mbox{$R = 0.4$} implemented in the FastJet software 
package~\cite{Cacciari:2011ma}. 
The measured energies of the jet candidates are corrected to the hadronic scale using the local cell signal weighting 
scheme~\cite{PERF-2012-01} and their transverse momentum has to exceed \mbox{25\,\GeV}.
For central jets ($|\eta| < 2.4$) with $\pt < \mbox{50\,\GeV}$, the scalar sum of the transverse momenta of tracks associated
with 
the jet and originating from the primary vertex of the interaction is required to be at least 50\% of the jet \pt.
This requirement suppresses jets originating from pile-up interactions~\cite{PERF-2014-03}.

The possible overlap between the object candidates is removed by applying the following requirements sequentially. 
Any electron that lies within a cone of size \mbox{$\Delta R = 0.1$} around a more energetic electron candidate or a muon 
candidate is discarded. 
Photon candidates are rejected if their angular distance to any remaining electron or muon is smaller than \mbox{$\Delta R = 
0.5$}. 
Apart from the removal of overlapping objects, this requirement also suppresses photons that are radiated from the lepton 
in the final state. 
Jets are discarded if they lie within a cone of size \mbox{$\Delta R = 0.3$} around an electron or \mbox{$\Delta R = 0.5$} around 
a photon candidate. 
Finally, muon candidates are rejected if their angular distance to a jet is smaller than \mbox{$\Delta R = 0.3$} in order to
remove muons originating from heavy-flavour quark decays within jets. 
 
The missing transverse momentum vector (\pet) of an event is a measure of the momentum imbalance in the transverse plane. 
It is calculated as the negative vector sum of the transverse momenta of calibrated leptons, photons, and jets, and additional 
tracks from the primary vertex that are not associated with any of those objects~\cite{PERF-2014-04}. 
The missing transverse momentum (\met) is defined as the magnitude of \pet.

The missing transverse momentum is employed for the definition of the selection criteria of the semileptonic analysis described 
in Section~\ref{sec:semi}. In the fully leptonic analysis, described in Section~\ref{sec:fully}, the relative missing transverse 
momentum (\metRel) is used as this improves the signal significance. Its definition is based on the absolute azimuthal separation
($\Delta\phi$) of the object closest to $\pet$: 

\begin{equation}
 \metRel = 
 \begin{cases}
    \met \times \sin(\Delta\phi),& \text{if } \Delta\phi(\pet,\,\text{closest object}) < \frac{\pi}{2},  \\
    \met                        ,& \text{otherwise}.
\end{cases}
\end{equation}

The transverse mass (\mt) is defined using \met, the transverse momentum ($p_{\text{T}}^{\ell}$) of the most energetic 
lepton in the event and the absolute angular difference between \pet and this lepton ($\Delta\phi (\pet, \ell)$):

\begin{equation}
\mt = \sqrt{2 p_{\text{T}}^{\ell} \met [1- \cos(\Delta\phi (\pet, \ell))]}.
\end{equation}

\section{Analysis of fully leptonic final states}
\label{sec:fully}
In the fully leptonic analysis, \wwy events are studied solely in the \emy final state.
Events where the two \Wb bosons decay to leptons of the same flavour, i.e. $e\nu e\nu \gamma$ or $\mu\nu\mu\nu\gamma$ final
states, have large backgrounds from Drell--Yan processes with photon radiation (\zy) and do not increase the sensitivity of this
measurement. 

The event selection for the fully leptonic analysis requires the presence of exactly one electron and one muon with opposite
electric charge, each with a transverse momentum of at least \mbox{20\,\GeV}, at least one reconstructed photon with $\et > 
\mbox{15\,\GeV}$, and relative missing transverse momentum larger than \mbox{15\,\GeV}. Events containing a third reconstructed 
electron or muon with $\pt > \mbox{7\,\GeV}$ are discarded to suppress backgrounds from $WW$ and $WZ$ diboson production. 
For the rejection of Drell--Yan background decaying to \mbox{$\tau$ leptons}, the invariant mass of the electron--muon pair is
required to be larger than \mbox{50\,\GeV}. 
Finally, events containing any reconstructed jet with $\pt > \mbox{25\,\GeV}$ are discarded, thereby reducing background
contributions from top-quark production. 
These selection requirements are optimised to yield the best sensitivity to the signal and define the signal region. 
The expected number of signal events is $12.2 \pm 1.1$, as computed with the VBFNLO program and corrected for
acceptance and efficiency effects (described in Section~\ref{sec:crosssec} along with the corresponding uncertainties).
A total of 26~events are observed.

\begin{table}[]
  \centering 
    \begin{tabular}{lr@{\,$\pm$\,}ll}
    \toprule
    \toprule
    Process  &   \multicolumn{2}{c}{Events} & Estimation Method\\
    \midrule
    \ttbary                   & 4.1 & 1.9 & MC simulation \\ 
    \zy                       & 2.7 & 1.2 & MC simulation \\
    $WZ\gamma$                      & 2.7 & 0.6 & MC simulation \\ 
    Fake $\gamma$ from $e$    & 2.3 & 0.6 & Corrected simulation\\ 
    Fake $\gamma$ from jets   & \multicolumn{2}{c}{$1.7\ ^{+\,3.3}_{-\,1.4}$} & 2D sideband method\\
    \wwy ($\tau$ contribution)& 1.0 & 0.1 & MC simulation \\ 
    $Wt$                      & 0.3 & 0.1 & MC simulation \\
    $ZZ$                      & 0.2 & 0.1 & MC simulation \\ 
    Fake $\mu$ from jets      & 0.1 & 0.1 & MC simulation \\ 
    Fake $e$ from jets        & \multicolumn{2}{c}{$0.0\ ^{+\,0.6}_{-\,0.0}$} & 2D sideband method\\
    \midrule
    Total background          & 15.1 & 4.1 & Sum of components \\ 
    Expected signal           & 12.2 & 1.1 & Corrected VBFNLO  \\ 
    Data                      & \multicolumn{2}{c}{26} & Measurement \\ 
    \bottomrule
    \bottomrule
  \end{tabular} 
  \caption{Expected and observed event yields for the fully leptonic final state in the \emy signal region. 
For each background process the corresponding estimation method is stated along with the resulting event yield. 
The quoted uncertainties include statistical and systematic uncertainties. 
The uncertainty in the total background expectation is symmetrised. 
The expected signal is computed with the VBFNLO program and corrected for acceptance and efficiency. 
} 
  \label{tab:fullyBGResults}
\end{table}

Several processes are backgrounds to the fully leptonic \wwy signal; their contributions in the signal region are summarised in 
Table~\ref{tab:fullyBGResults}. 
The dominant source of background is the production of $t\bar{t}\gamma$ events where the top quarks decay to \Wb bosons and
$b$-quarks with a leptonic decay of the \Wb boson ($t \rightarrow Wb \rightarrow \ell \nu b$).
This process mimics the signal when the jets have low energy or are produced in the forward direction ($|y| \geq 4.4$) and hence
the jets are not reconstructed. 
Other subdominant backgrounds are \zy events, which contribute when the \Zb boson decays to a pair of leptonically 
decaying $\tau$ leptons, and $WZ\gamma$ production, which can mimic the signal when one of the final state leptons does not 
fulfil the identification criteria or is not reconstructed due to the limited geometrical acceptance.
Other backgrounds arise from \wwy production including \mbox{$\tau$ leptons} and the production of $Wt$ and $ZZ$ events.
The event yields of all these processes are estimated using MC simulation.
The corresponding uncertainties include statistical and systematic uncertainties that are of similar size.
The systematic uncertainties can be subdivided into experimental uncertainties and uncertainties from the theoretical
calculation. The two components contribute equally to the uncertainty for most processes. The relative uncertainties from the
theoretical calculation range from 5\% to 22\%~\cite{PhysRevD.83.074013, STDM-2014-01, Campbell:2011bn, 
Campanario:2012fk, Cascioli:2014yka, TOPQ-2012-20}; 
the uncertainties associated with the computation of the \wvy process are described in Section~\ref{sec:crosssec}.
The experimental uncertainties include the energy scale and energy resolution uncertainties of the reconstructed
objects~\cite{PERF-2013-05, PERF-2014-05, PERF-2012-01, PERF-2011-04, PERF-2011-07}, 
the uncertainties associated with the efficiencies of their reconstruction and
identification~\cite{PERF-2016-01,ATL-PHYS-PUB-2011-007, PERF-2014-05}, 
as well as uncertainties attributed to the simulation of the event pile-up~\cite{PERF-2014-03}. 
The relative experimental uncertainties range from 5\% to 32\% with the largest contribution arising from the jet energy
scale uncertainty which mainly contributes due to the requirement that the signal events should not contain reconstructed jets.

Events containing misidentified objects also constitute an important source of background.
The background from $WZ$ production where an electron is reconstructed as a photon (fake $\gamma$ from $e$) is
estimated by using MC simulation, where the rate of electrons being reconstructed as photons is corrected to better describe the 
data. 
This rate is determined by studying the decays of \Zb bosons to two electrons where one of the electrons is reconstructed as a 
photon and is below 6\% for most of the pseudorapidity region.
The uncertainty of this correction is small compared to the total uncertainty, which also includes the statistical 
uncertainty, uncertainties from the theoretical calculation, and experimental uncertainties as discussed in the previous 
paragraph.

\begin{figure}[]
 \centering
 \includegraphics[width=.7\textwidth]{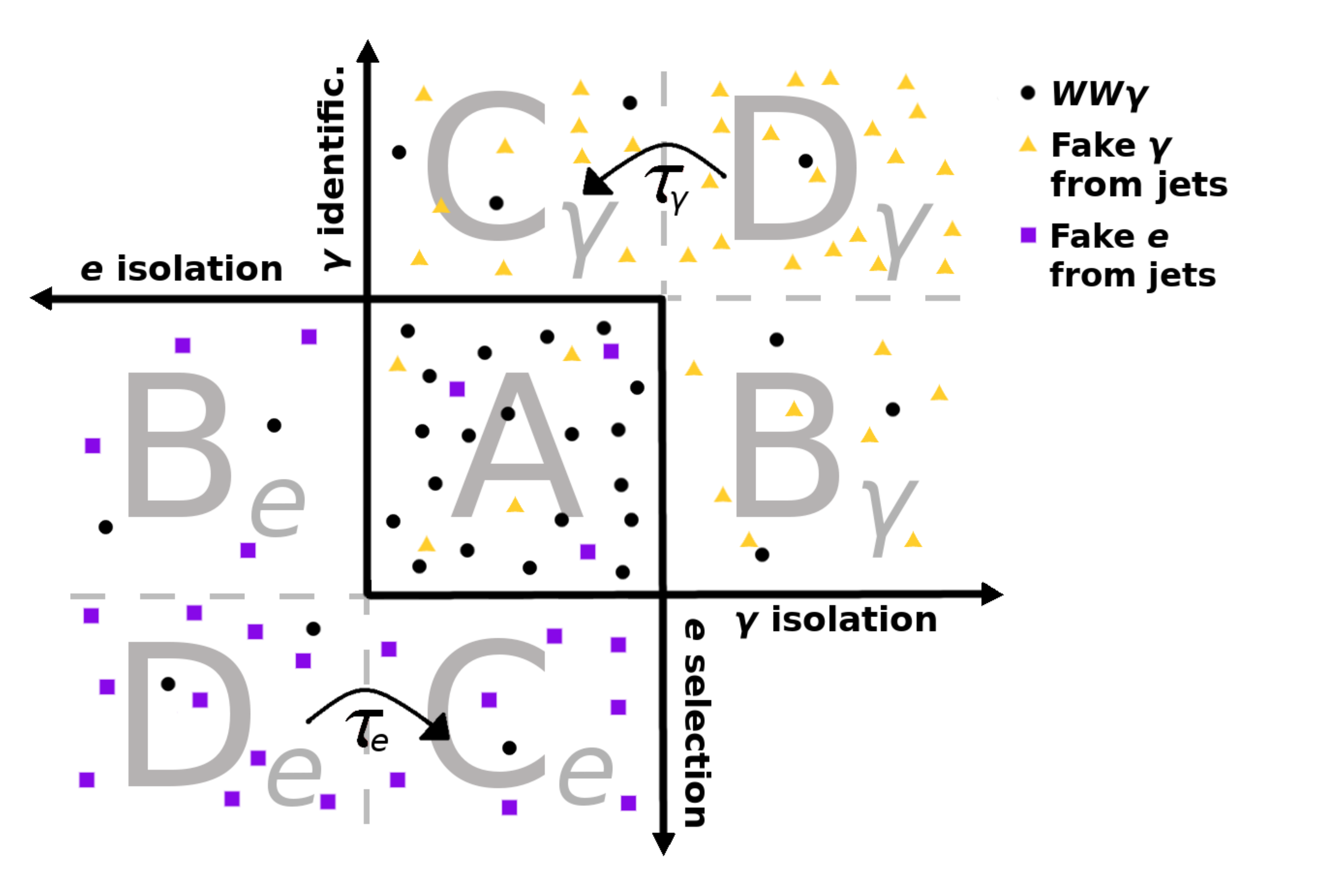}
 \caption{Schematic drawing of the combination of the two 2D sideband methods to estimate the background from events containing
fake $\gamma$ (triangles) and fake $e$ (squares) from jets. The \wwy events are indicated with filled circles. 
The figure shows the signal region (region A) along with the six sideband regions. In regions C$_{\gamma}$ and D$_{\gamma}$ the
requirement on the electron isolation stays unchanged as does the requirement on the photon isolation in regions C$_{e}$ and
D$_{e}$. 
The factors $\tau_{\gamma}$ and $\tau_{e}$ that relate the event count in the isolated and non-isolated fake-object regions are
also shown. 
The contributions of SM background processes to the different regions are omitted for simplicity.}
  \label{fig:abcdMethodComb}
\end{figure}

The production of $WW$ and $t\bar{t}$ pairs in association with jets can mimic the signal if jets are misidentified as photons
(fake $\gamma$ from jets). Jets can also be misidentified as muons (fake $\mu$ from jets) or electrons (fake $e$ from jets) in
which case \wy+ jets events can fulfil the signal selection criteria.
The contribution from events containing fake $\mu$ from jets is determined from MC simulations and found to be very small. Events
including fake $\gamma$ from jets or fake $e$ from jets are removed from the MC simulation, as their contribution is 
estimated with data. 
These contributions are estimated by combining two two-dimensional (2D) sideband methods~\cite{STDM-2012-07} (one per
background component). 
A schematical drawing of the interplay between the methods is given in Figure~\ref{fig:abcdMethodComb}. 
It shows the three background-enriched sideband regions (B$_x$, C$_x$, D$_x$) per fake-object category $x$ (with $x \in \{\gamma, 
e\}$) along with the signal region (A) that is common to the two fake-object categories. 
In the sideband regions, the contribution from signal and other SM processes containing prompt photons is accounted for using MC 
estimates.
The method relies on the assumption that the definition of the sideband regions uses uncorrelated observables. 
Then, the ratio $\tau_{x}$ of the number of events in region C$_x$ ($N^{\text{fake}\,x}_{\text{C}_{x}}$)
to the number of events in region D$_x$ ($N^{\text{fake}\,x}_{\text{D}_{x}}$) multiplied by the number of events in region
B$_x$ ($N^{\text{fake}\,x}_{\text{B}_{x}}$) can be used to estimate the number of events containing fake objects of category $x$
in region A ($N^{\text{fake}\,x}_{\text{A}}$).
A possible correlation of the observables is accounted for by introducing the correlation factor $\rho_{x}$, which is set to one, 
representing no correlation, for the computation of the background contributions and varied to estimate the corresponding 
uncertainty. 

The sideband regions B$_{\gamma}$, C$_{\gamma}$ and D$_{\gamma}$ are defined using the photon isolation, \isoy, and a 
set of photon identification criteria related to the energy deposits in the first layer of the LAr calorimeter.
The sideband regions B$_{e}$, C$_{e}$ and D$_{e}$ are defined using the electron isolation, \isoe, and a set of 
electron--jet event selection criteria. 
The latter require the presence of at least one candidate electron and one jet with an absolute azimuthal separation of at least 
0.7 in the event as well as $\mt \leq \mbox{30\,\GeV}$ and, if there is a second lepton in the event, the invariant mass of the 
lepton pair, \mll, has to fulfil\footnote{The mass of the \Zb boson is taken to be $m_{Z} = 
\mbox{91.19\,\GeV}$~\cite{Olive:2016xmw}.} $|\mll - m_{Z}| > \mbox{7\,\GeV}$. 
The latter two criteria suppress the contribution of electrons originating from the decay of \Wb and \Zb bosons, respectively.

As region A is common to the two fake-object categories, the estimation of the fake $\gamma$ and fake $e$ from jets 
contributions in the signal region is performed simultaneously using a maximum likelihood approach.
The likelihood function is the product of the Poisson probabilities of observing the expected number of events in the seven
regions multiplied by Gaussian functions that incorporate the systematic uncertainties as nuisance parameters.
This function has seven free parameters: 
the number of signal events in the signal region ($N^{\emy}_{\text{obs}}$), the ratios $\tau_{\gamma}$ and $\tau_{e}$ as well as 
$N^{\text{fake}\,\gamma}_{\text{A}}$, $N^{\text{fake}\,e\vphantom{\gamma}}_{\text{A}}$,
$N^{\text{fake}\,\gamma}_{\text{C}_{\gamma}}$ and $N^{\text{fake}\,e\vphantom{\gamma}}_{\text{C}_{e\vphantom{\gamma}}}$.
These parameters are determined by maximising the likelihood function that is constrained using the number of observed events in 
the seven regions defined by the method.

Apart from providing the contribution of fake $\gamma$ and fake $e$ from jets in the signal region, the likelihood function
also yields the most likely value of the number of signal events in the signal region: \mbox{$N^{\emy}_{\text{obs}} = 9.4 \pm
6.2$}.
This value is consistent with the difference between the number of observed events and the total background prediction given in 
Table~\ref{tab:fullyBGResults}. 
The former is used for the determination of the fiducial cross-section in Section~\ref{sec:crosssec}. 
Several sources of systematic uncertainty are taken into account. 
Varying the correlation factor $\rho_{\gamma}$ ($\rho_{e}$) from one by its uncertainty $\Delta\rho_{\gamma}^{MC} = \pm \ 0.44$ ($\Delta\rho_{e}^{MC} = \pm \ 0.69$) as extracted from the MC simulation expectation, yields a relative uncertainty in $N^{\emy}_{\text{obs}}$ of 10\% (0.4\%).
The uncertainty in the number of events from SM processes in the sideband regions that are estimated from simulation is
accounted for by varying the event yield by its total uncertainty and contributes 6\% to the total uncertainty in 
$N^{\emy}_{\text{obs}}$.
The uncertainty in estimating the number of signal events in the sideband regions contributes less than 1\% to the total 
uncertainty.
The dominant uncertainty in $N^{\emy}_{\text{obs}}$ originates from the limited number of data events and contributes a relative 
uncertainty of 60\%.

\begin{figure}[]
 \centering
 \includegraphics[width=.7\textwidth]{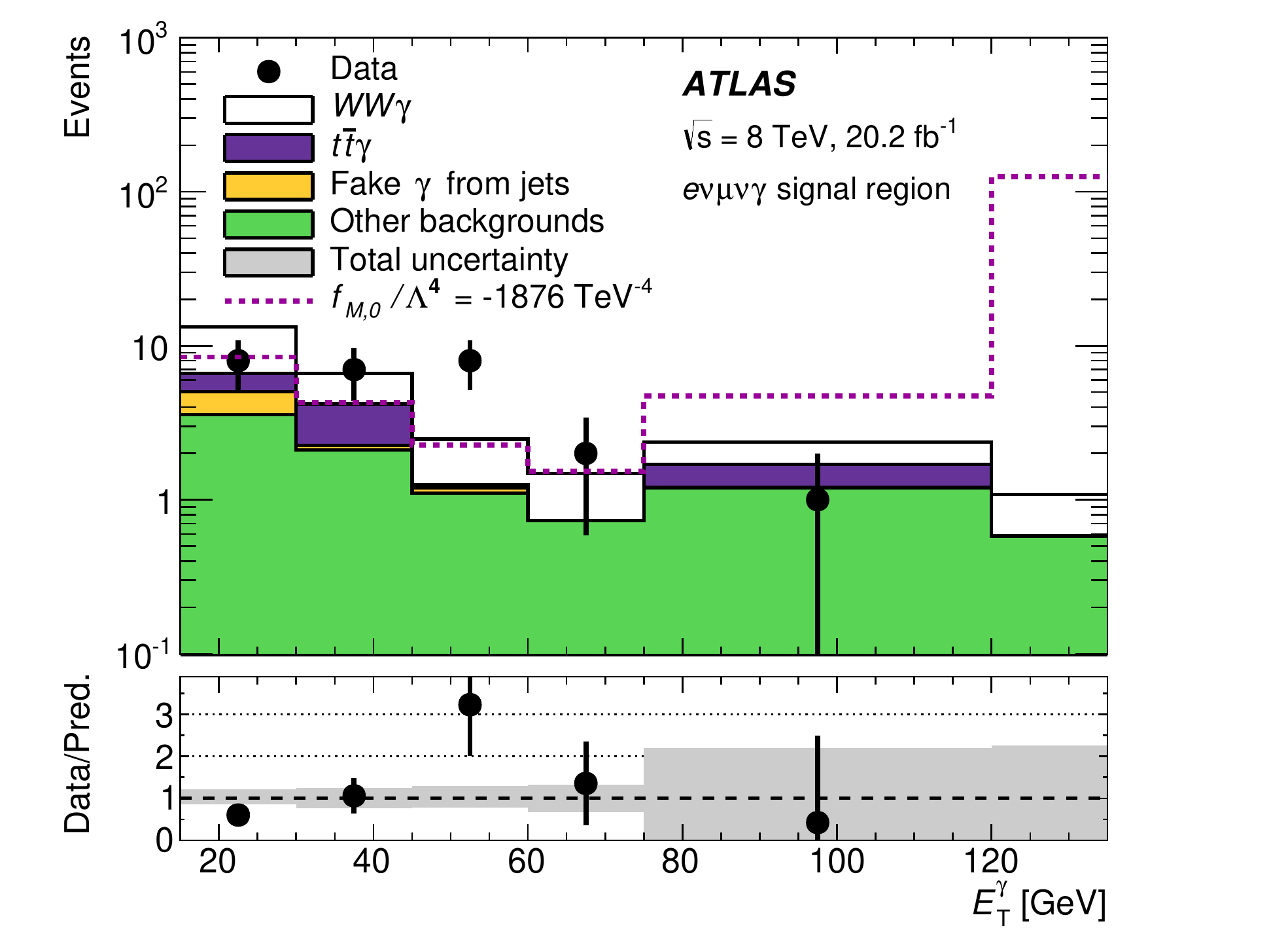}\label{fig:ph_pt}
 \caption{Observed and expected transverse energy distribution of the photon with the highest \et in the \emy signal region. 
The data are shown together with the predicted signal and backgrounds.
Also indicated is the expected event yield for a reference model describing aQGCs with \mbox{$f_{M,0}/\Lambda^{4} =
-1876\,\TeV^{-4}$} (see Section~\ref{sec:aqgc}).
The last bin contains all overflow events.
The lower panel shows the ratio of the observed number of events to the sum of expected signal and background events as well
as the corresponding uncertainties.}
 \label{fig:fullyDistr}
\end{figure}
 
Figure~\ref{fig:fullyDistr} shows the transverse energy distribution of the photon with the highest \et in the signal region.
The data are shown together with the expected signal from the MC prediction and the results from the background estimation. 
Also shown is the predicted event yield for a reference point in the parameter space of aQGCs discussed in Section~\ref{sec:aqgc}. 
The lower panel of the figure shows the ratio of the number of observed events to the sum of the expected
signal and background events.

\section{Analysis of semileptonic final states}
\label{sec:semi}
In the semileptonic analysis, \wvy production with one leptonically decaying \Wb boson and one hadronically decaying \Wb or \Zb
boson is studied. 
The event selection requires one lepton, at least two jets, at least one photon, and missing transverse momentum. 
The analysis is performed separately in the electron and the muon channels. 
The transverse momentum of the reconstructed electron or muon is required to be larger than \mbox{25\,\GeV}. 
Events containing additional reconstructed electrons or muons with \mbox{$\pt > \mbox{7\,\GeV}$} are discarded. 
Photons are required to have $\et > \mbox{15\,\GeV}$. 
Jets are required to have $\pt > \mbox{25\,\GeV}$ and to be within the volume of the tracking detector, \mbox{$|\eta| < 2.5$}, to
ensure that jets originating from heavy-flavour quarks can be identified. 
In addition, the two jets with the highest transverse momenta are required to be close together with \mbox{$|\Delta \eta_{jj}| < 
1.2$} and \mbox{$\Delta R_{jj} < 3.0$} to reject backgrounds from \mbox{\wg+ jets} events.
The missing transverse momentum and the transverse mass of the event are both required to exceed \mbox{30\,\GeV}.
In events containing electrons, the invariant mass of the electron--photon pair is required to differ from the value of the \Zb 
boson mass by at least \mbox{10\,\GeV} to suppress backgrounds from events containing leptonically decaying \Zb bosons. 
To reduce background contributions from processes including top quarks, mainly \ttbary, events containing jets that are identified
as originating from the decay of a $b$-hadron are rejected. 
The $b$-jet identification is performed using the MV1 algorithm~\cite{PERF-2012-04} based on an artificial neural network with
an efficiency of 85\% and a light-quark-jet and gluon-jet misidentification rate of 10\%. 
Finally, the invariant mass of the two jets with the highest transverse momenta in the event is required to be close to the mass
of the decaying \Wb or \Zb boson, i.e. \mbox{$\mbox{70\,\GeV} < m_{jj} < \mbox{100\,\GeV}$}. 
These selection requirements are optimised to yield the best sensitivity to the signal and define the signal region. 
The expected number of signal events is $14 \pm 2$ ($18 \pm 2$) in the electron (muon) channel, as computed with the
VBFNLO program and corrected for acceptance and efficiency effects (described in Section~\ref{sec:crosssec} along with the
corresponding uncertainties).
A total of 490~(599)~events are observed in the electron (muon) channel.

\begin{table}[]
\centering
\sisetup{round-mode = places}
\sisetup{round-precision = 0}
 \begin{tabular}{lr@{\,$\pm$\,}lr@{\,$\pm$\,}ll}  
  \toprule
  \toprule
  Process                 & \multicolumn{2}{c}{Electron Channel}    & \multicolumn{2}{c}{Muon Channel} & Estimation Method\\
  \midrule
  \wy+ jets               & \num{323.996} &            \num{11.047} & \num{406.934} & \num{11.397 }   & Simultaneous fit\\
  Fake $\gamma$ from jets & \num{ 82.457} & \phantom{1}\num{ 7.199} & \num{117.275} & \phantom{1}\num{ 8.793 }& Simultaneous
fit  \\
  Fake $\ell$ from jets   & \num{ 56.948} & \phantom{1}\num{ 5.735} & \num{ 26.935} & \phantom{1}\num{ 5.481 }& Simultaneous
fit \\
  \ttbary                 & \num{ 35.350} & \phantom{1}\num{ 5.778} & \num{ 46.342} & \phantom{1}\num{ 7.255 }& MC simulation \\
  Fake $\gamma$ from $e$  & \num{ 33.150} &            \num{12.272} & \num{  2.872} & \phantom{1}\num{ 0.773 }& Corrected
simulation \\
  \zy+ jets               & \num{ 19.367} & \phantom{1}\num{ 3.730} & \num{ 19.938} & \phantom{1}\num{ 2.863 }& MC simulation \\
  \wvy ($\tau$ contribution)& \multicolumn{2}{c}{$< 1$} \phantom{111}& \multicolumn{2}{c}{$< 1$} \phantom{.}  & MC simulation \\
  \midrule                                                   
  Total background        & \num{552} &            \num{38}   & \num{621} &  \num{31 } & Sum of components \\  
  Expected signal         & \num{ 14 } &            \phantom{1}\num{2}   & \num{ 18 } &  \phantom{1}\num{ 2} & 
Corrected VBFNLO\\         
  Data                    & \multicolumn{2}{c}{490} \phantom{111} & \multicolumn{2}{c}{599} \phantom{.} & Measurement\\   
  \bottomrule
  \bottomrule
 \end{tabular} 
 \caption{Expected and observed event yields in the signal region of the electron and muon channels of the semileptonic analysis.
For each background process the corresponding estimation method is stated. The uncertainties of the \mbox{\wg+ jets}, fake
$\gamma$ from jets and fake $\ell$ from jets are solely the statistical uncertainties from data. The uncertainties of the \ttbary,
fake $\gamma$ from $e$, \mbox{\zy+ jets} and $\wvy \rightarrow \tau \nu jj \gamma$ backgrounds correspond to the sum in 
quadrature of the statistical uncertainty of the MC simulation and the uncertainties of the theoretical prediction. 
The uncertainty in the total background estimate is symmetrised and contains the statistical uncertainty of the data, the 
uncertainties of the theoretical prediction, and experimental uncertainties. 
The expected signals are computed with the VBFNLO program and corrected for acceptance and efficiency.}
\label{tab:SimSR_numbers}
\end{table}

The background processes of the semileptonic analysis are listed in Table~\ref{tab:SimSR_numbers}. 
The dominant contribution arises from \mbox{\wg+ jets} production, as it has the same final state as the signal. 
The contribution from \ttbarg, \mbox{\zy+ jets} as well as from \wvy processes containing $\tau$ leptons ($\wvy \rightarrow 
\tau \nu jj \gamma$) processes, is estimated using MC simulation. 
The uncertainties in these background contributions given in Table~\ref{tab:SimSR_numbers} solely include statistical
uncertainties and the uncertainties of the theoretical prediction, that are of the same size. 
The relative uncertainties of the theoretical predictions range from 4\% to 22\%~\cite{PhysRevD.83.074013, STDM-2014-01};
the uncertainties associated with the computation of the \wvy process are described in Section~\ref{sec:crosssec}. 
The experimental uncertainties are only included in the uncertainty of the total background estimation in
Table~\ref{tab:SimSR_numbers}, as they are correlated for the individual background components. 

Events containing misidentified objects constitute an important source of background in this analysis as well. 
When electrons are misidentified as photons (fake $\gamma$ from $e$), $Z \rightarrow ee$ production in association with jets and
$t\bar{t}$ events can mimic the signal. 
As in the fully leptonic analysis, this background is estimated using MC simulation which is corrected to match the
misidentification rate measured in data. 
The uncertainty of this correction is small compared to the statistical uncertainty and the uncertainties from the 
theoretical calculation. The latter uncertainty is estimated to be 5\% for the $Z \rightarrow ee$ and the $t\bar{t}$
processes in agreement with the corresponding measurements~\cite{STDM-2014-12, TOPQ-2015-07}.
Mainly events from \mbox{$W \ +$ jets} production contribute as background when a jet is misidentified as a photon (fake $\gamma$
from jets).
In events containing jets misidentified as leptons (fake $\ell$ from jets) predominantly \mbox{$\gamma \ +$ jets} production
constitues a background. 
Events containing fake $\gamma$ from jets or fake $\ell$ from jets are removed from the MC simulation, as their contribution is
estimated with data.

A simultaneous fit is used to estimate the background contributions from \mbox{\wg+ jets} production and from events containing
fake $\gamma$ from jets and fake $\ell$ from jets (the fake $e$ from jets component also includes the small contribution from fake
$e$ from $\gamma$).
The simultaneous fit consists of three components: 
a binned extended maximum-likelihood fit of the invariant dijet mass distribution to constrain the \mbox{\wg+ jets} 
contribution, a binned extended maximum-likelihood fit of the \met distribution to constrain the fake $\ell$ backgrounds and a 
two-dimensional sideband method to constrain the contribution from fake $\gamma$ from jets. 
The free parameters of the simultaneous fit are the normalisation of the \mbox{\wg+ jets} background, the normalisation of the
processes containing fake $\ell$ from jets and the normalisation of the processes containing fake $\gamma$ from jets. 
The normalisation of all other background components is fixed.
The fit is performed separately in the electron and muon channels of the analysis. 
For all three estimation methods the signal region with \mbox{$\mbox{70\,\GeV} < \mjj < \mbox{100\,\GeV}$} is excluded such that
the overall signal contribution to the fiducial region used for the background estimation is negligible. 
Therefore, the signal contribution in all regions used in the fit is neglected and the result is independent of the signal
modelling.
The \mjj distribution is fitted in the range \mbox{10--70\,\GeV} and \mbox{100--505\,\GeV}; the \met distribution is fitted in
the range \mbox{0--300\,\GeV}. 
No minimum \met requirement is imposed in the fit of the \met distribution, in order to increase the sensitivity to fake
$\ell$ from jets, as these events are expected to have low missing transverse momentum. 
Apart from neglecting the signal contribution, the two-dimensional sideband method is performed as for the fake photons from jets
in the fully leptonic analysis.

The extended likelihood fits employ shape templates for the \mjj and \met distributions of the different background components. 
The shape templates for all backgrounds are derived from simulation apart from the ones associated with fake $\ell$
from jets and fake $\gamma$ from jets. 
The latter shape templates are obtained from data events selected similarly to the fit regions with some requirements modified
as follows to enhance the contribution from the respective fake object. 
To estimate the shape template for fake $e$ from jets, the requirement on \met is removed and the requirements on the electron
identification and isolation are modified. 
To this end, the requirements on the calorimeter-based isolation and the origin of the electron track are removed and the 
track-based isolation requirement is inverted.
To estimate the shape template for fake $\mu$ from jets, the requirement on \met is removed and the requirements on the muon
isolation and the origin of the track are inverted. 
To estimate the shape template for fake $\gamma$ from jets, the requirement on the photon isolation is removed and at least one of
the photon identification criteria based on the energy deposits in the first layer of the LAr calorimeter must not be satisfied. 
The \mjj shape templates are also employed to extrapolate the background estimation results of the different background components
to the signal region. 

\begin{figure}[]
 \centering
 \subfloat[\met distribution - Electron
channel]{\includegraphics[width=.5\textwidth]{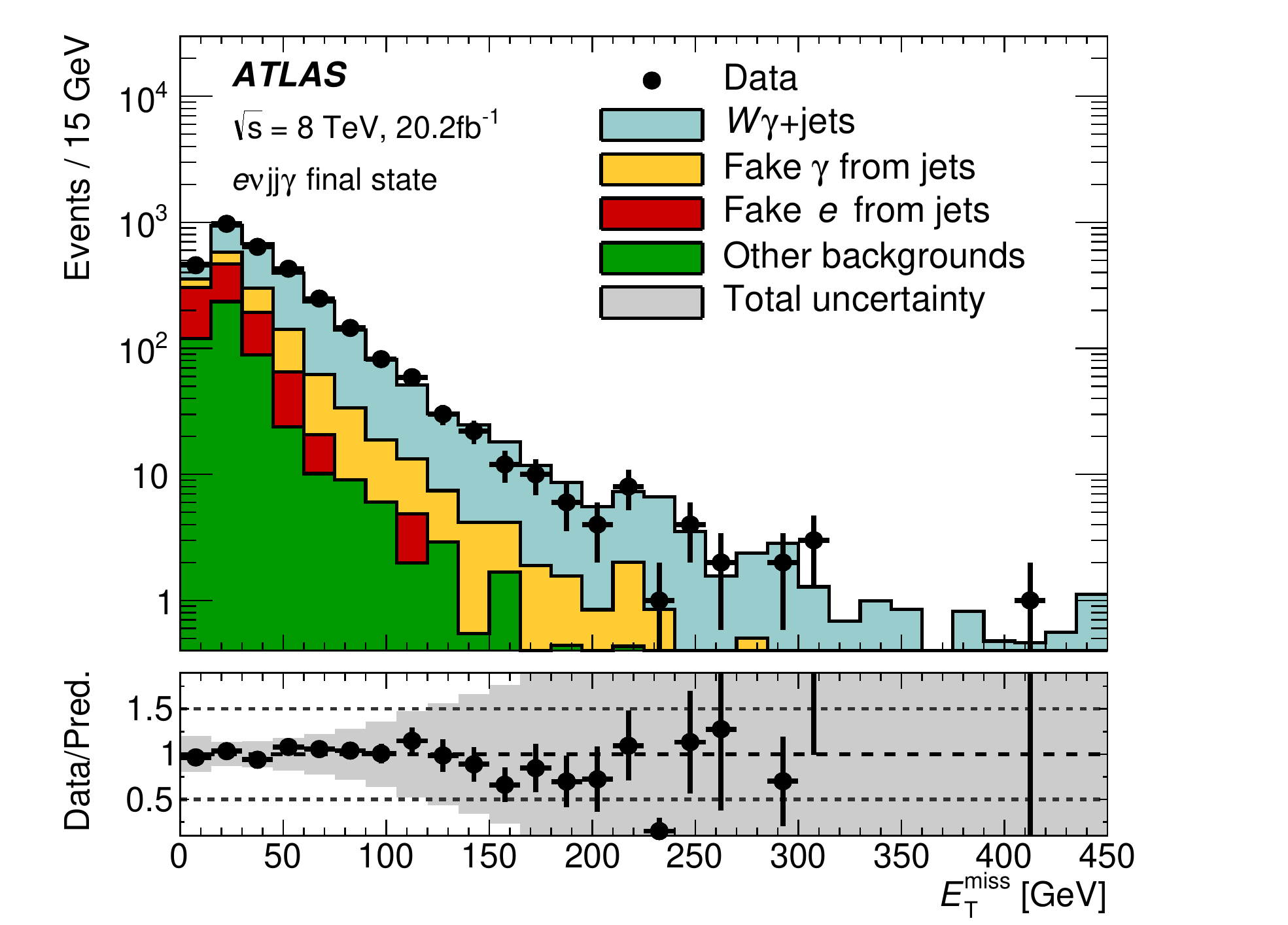}\label{fig:met_el}} 
\subfloat[\mjj distribution- Electron
channel]{\includegraphics[width=.5\textwidth]{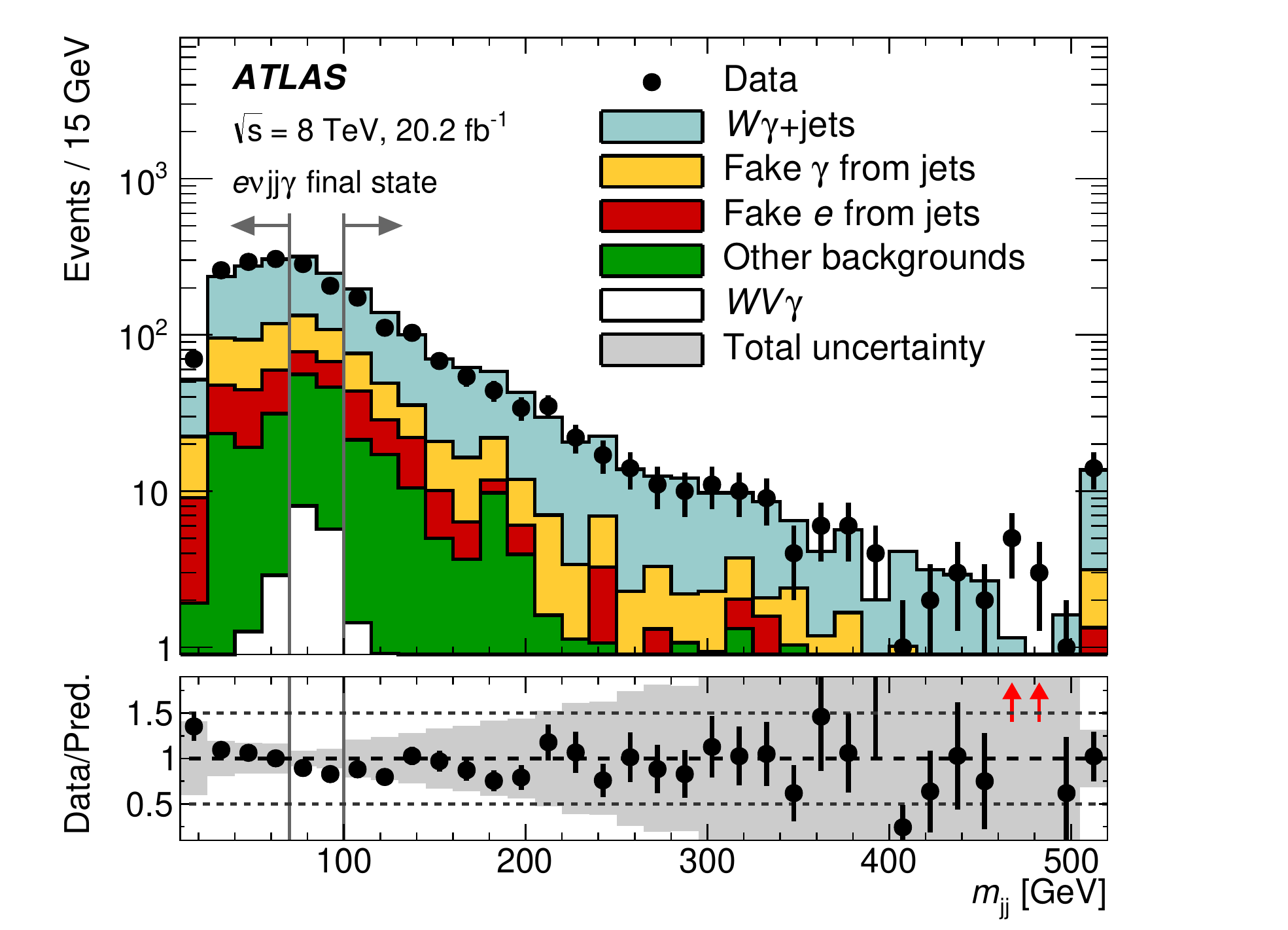}\label{fig:mjj_el}} \\
 \subfloat[\met distribution - Muon
channel]{\includegraphics[width=.5\textwidth]{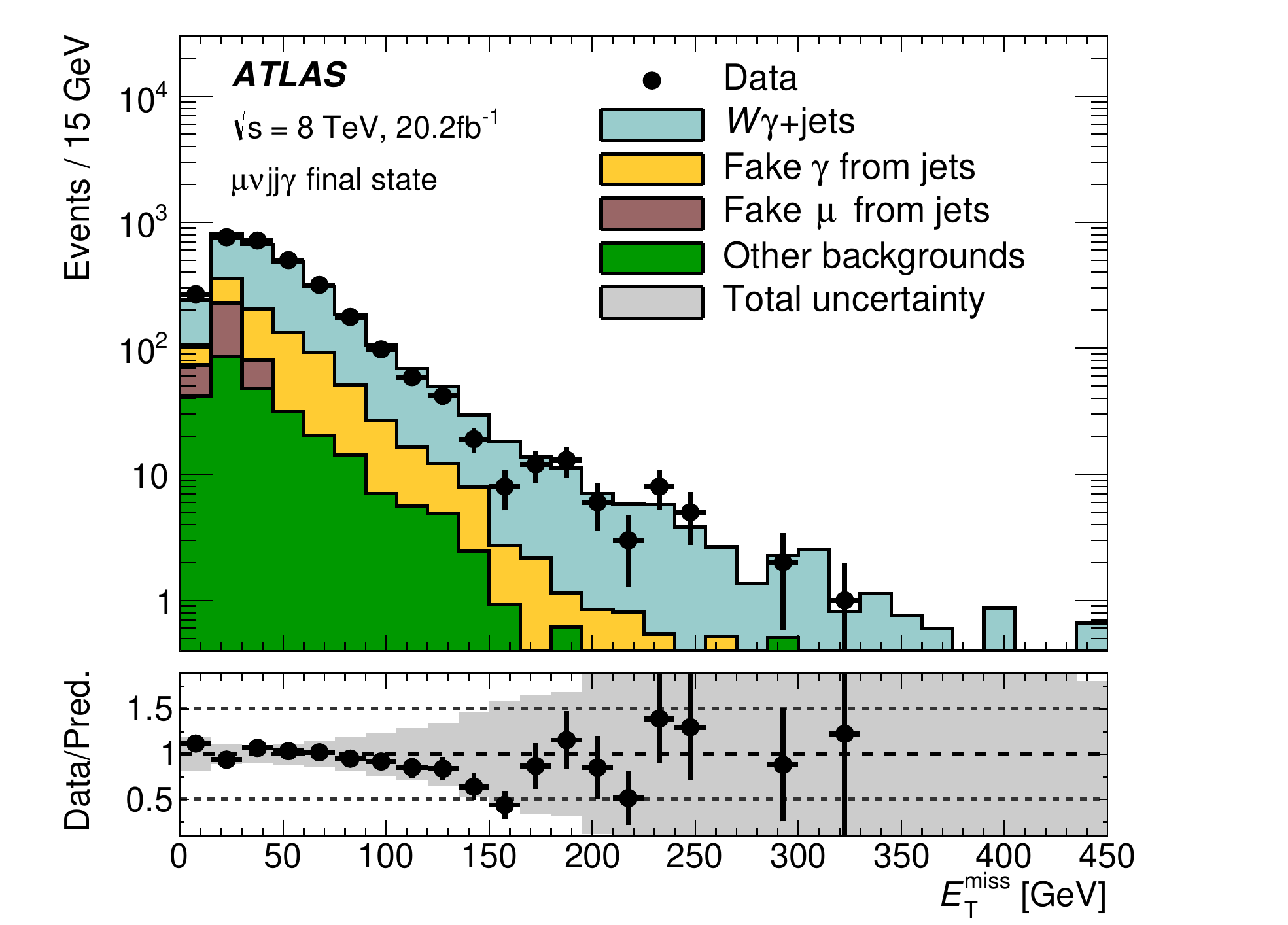}\label{fig:met_mu}}
 \subfloat[\mjj distribution - Muon
channel]{\includegraphics[width=.5\textwidth]{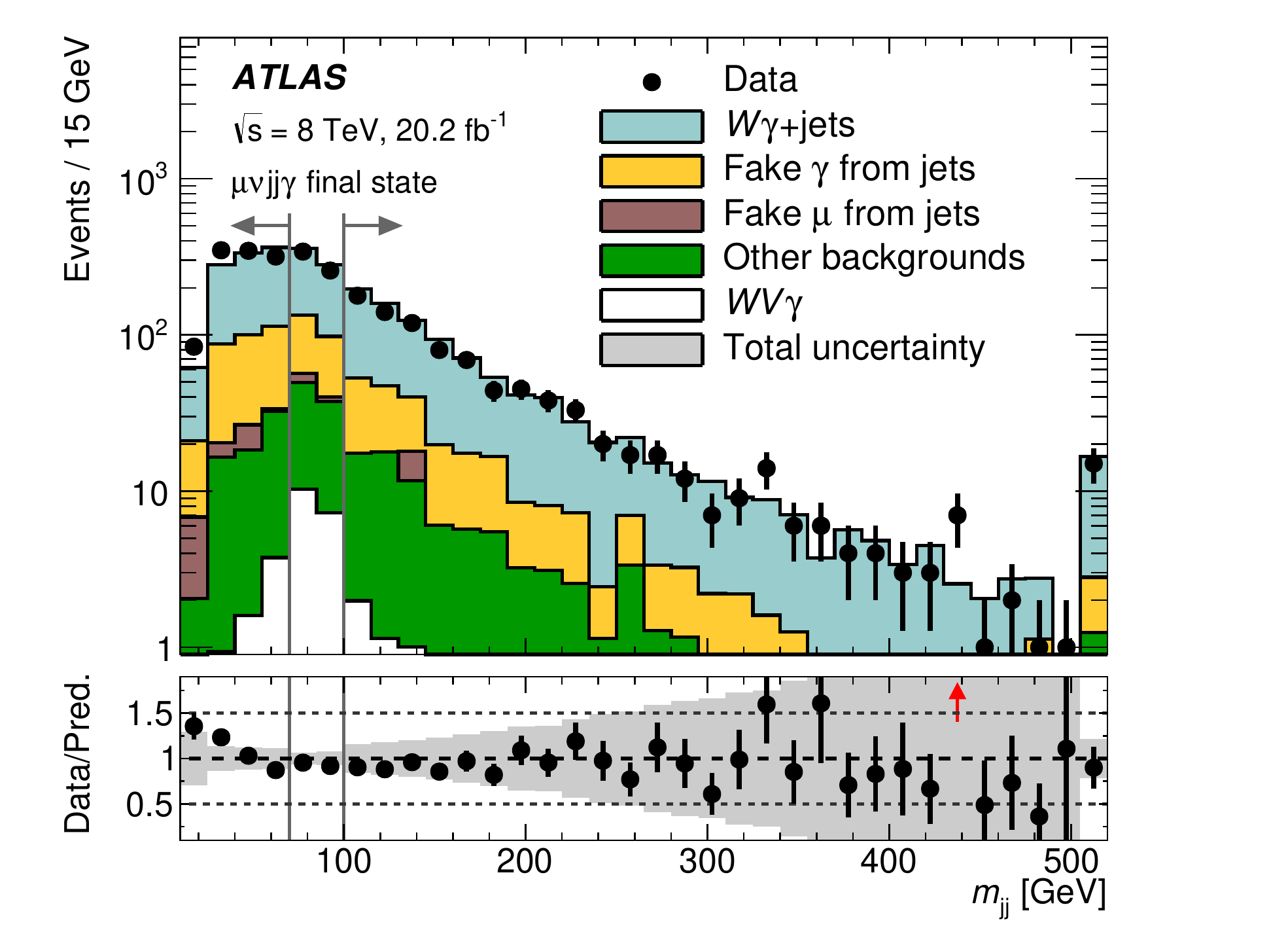}\label{fig:mjj_mu}}
 \caption{Missing transverse momentum and dijet invariant mass distributions of the electron (upper row) and the muon channels
(lower row) of the semileptonic analysis. The different background components are shown together with the data. 
The signal region (\mbox{$\mbox{70\,\GeV} < m_{jj} < \mbox{100\,\GeV}$}) is excluded
in~\protect\subref{fig:met_el} and~\protect\subref{fig:met_mu} as well as in the simultaneous fit as indicated by the arrows
in~\protect\subref{fig:mjj_el} and~\protect\subref{fig:mjj_mu}. 
The last bin of each figure contains the event overflow. 
The lower panels show the ratio of the observed number of events to the predicted background as well as the
corresponding uncertainties. 
The red arrows indicate entries that are outside the $y$-axis range.}
 \label{fig:muChFits}
\end{figure}

Figure~\ref{fig:muChFits} shows the results of the simultaneous fit, in the upper panel for the electron channel and in the 
lower panel for the muon channel. 
In Figures~\ref{fig:met_el} and~\ref{fig:met_mu} the resulting \met distributions are presented; the events are selected using 
the criteria for the signal region, but the requirement on \met is removed and the requirement on \mjj is inverted. 
The lower panels of the figures show the ratio of the observed number of events to the expected number of events, which agrees 
with unity within uncertainties. 
In Figures~\ref{fig:mjj_el} and~\ref{fig:mjj_mu}  the resulting \mjj distributions are shown. 
All signal selection requirements apart from the \mjj requirement are imposed. 
The distribution observed in data is underestimated by the background estimation in both channels at low \mjj values but agrees
within uncertainties. As a cross check, an alternative shape template for the \mbox{\wg+ jets} background is obtained from  
simulated events generated with SHERPA. 
While the resulting background estimate shows better agreement with the data at low values of \mjj, no significant impact on
the background estimate in the signal region is found. The event yields of the \mbox{\wg+ jets}, fake $\gamma$ from jets and 
fake $\ell$ from jets events in the signal region are given in Table~\ref{tab:SimSR_numbers}. The uncertainties in these 
components in Table~\ref{tab:SimSR_numbers} correspond solely to the statistical uncertainty from data. 

The uncertainty in the total number of background events has several sources.
The uncertainty associated with the shape templates is estimated by performing \mbox{10000\,pseudo experiments} that use
alternative
shape templates obtained from sampling the nominal ones bin-wise using a Gaussian distribution. 
The width of the Gaussian distribution corresponds to the statistical uncertainty of the shape templates determined from data, or
to the statistical uncertainty of the MC simulation and the uncertainties from the theoretical calculation if they are determined
from simulation.
The shape templates are varied simultaneously and yield an uncertainty in the total background of 5\% (4\%) in the
electron (muon) channel.
The experimental uncertainties are the uncertainties due to reconstruction and identification efficiencies of the
objects~\cite{PERF-2016-01,ATL-PHYS-PUB-2011-007, PERF-2014-05, ATLAS-CONF-2014-046, ATLAS-CONF-2014-004} including energy scale
and energy resolution uncertainties~\cite{PERF-2013-05, PERF-2014-05, PERF-2012-01, PERF-2011-04, PERF-2011-07} as well as
uncertainties arising from the simulation of the event pile-up~\cite{PERF-2014-03}. 
These uncertainties are estimated for all background components simultaneously and amount to a total of 4\% (3\%)
in the electron (muon) channel. They are dominated by the uncertainty in the jet energy scale.
The uncertainty related to the choice of fit boundaries for the extended maximum-likelihood fits is estimated by varying
these boundaries. The lower \mjj (\met) boundary is set to \mbox{25\,\GeV} (\mbox{15\,\GeV}) and the upper boundary is set to
\mbox{490\,\GeV} or \mbox{520\,\GeV} (\mbox{285\,\GeV} or \mbox{315\,\GeV}) independently. 
The uncertainty introduced by the choice of binning for the distributions used for the extended maximum-likelihood fits is
estimated by varying the bin sizes by a factor of two. 
The uncertainty due to the possible correlation of the selection criteria defining the sideband regions of the 2D sideband method is estimated by changing the value of the correlation factor $\rho$ from one by its uncertainty $\Delta\rho_{e\nu jj \gamma}^{MC} = \pm \ 0.38$ ($\Delta\rho_{\mu\nu jj \gamma}^{MC} = \pm \ 0.23$) as extracted from the MC simulation expectation.
The uncertainty associated with any of these fit parameter variations is less than 1\% in each channel of the analysis.
The statistical uncertainty in the expected total number of background events corresponds to 2.6\% (2.5\%) 
in the electron (muon) channel.  

\begin{figure}[]
 \centering
 \subfloat[Electron channel]{\includegraphics[width=.5\textwidth]{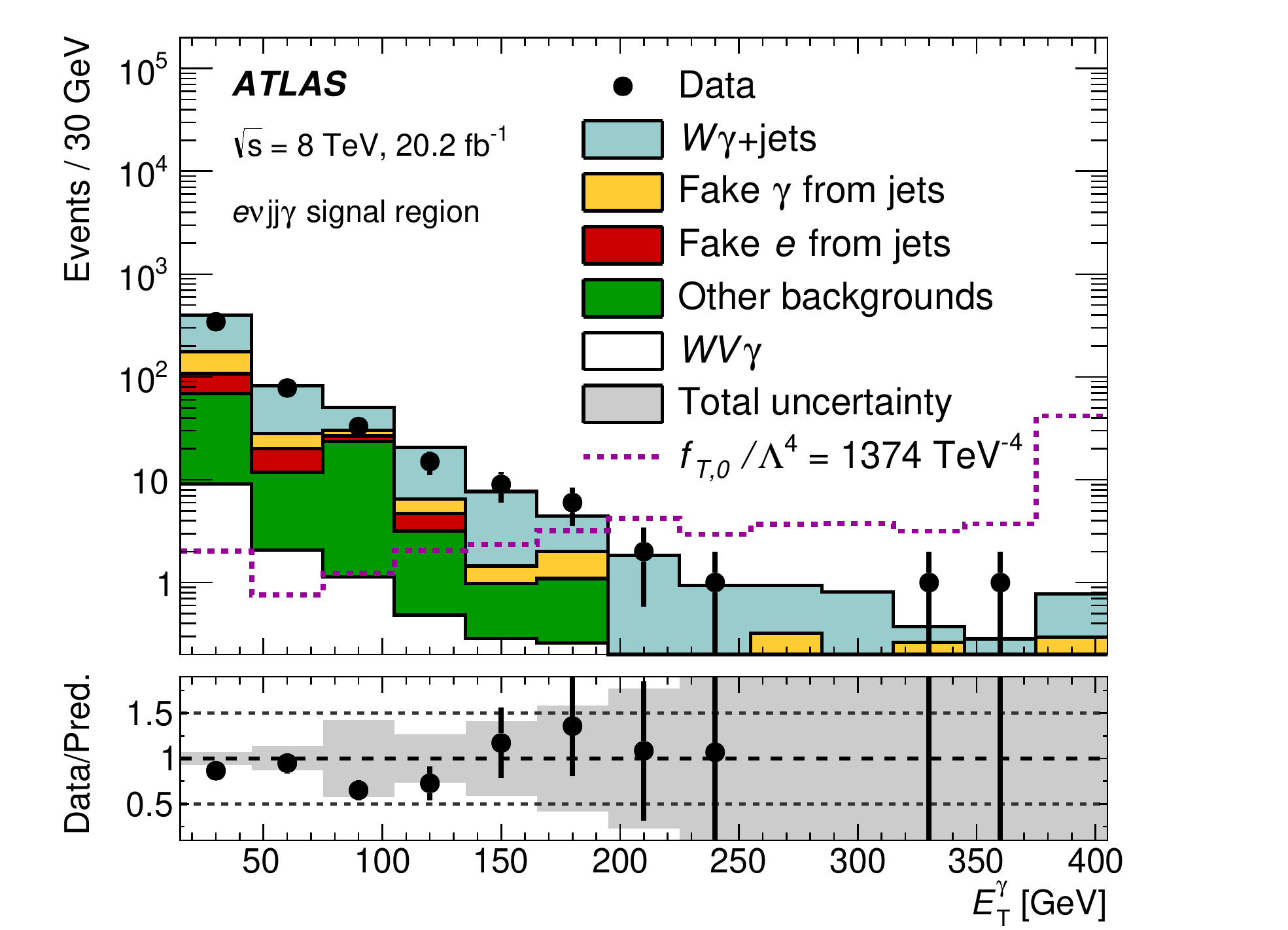}\label{fig:pTpho_el}}
 \subfloat[Muon channel]{\includegraphics[width=.5\textwidth]{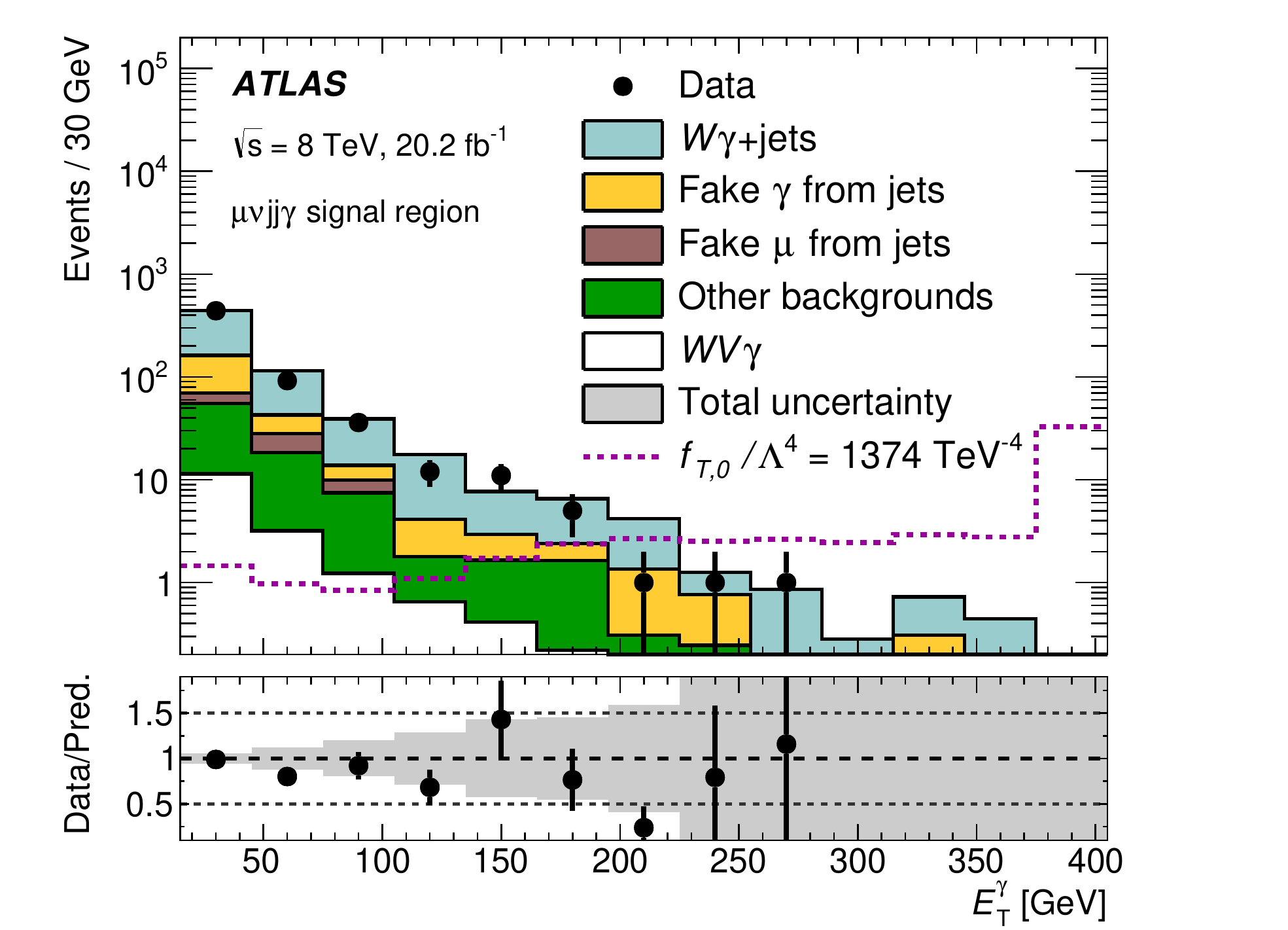}\label{fig:pTpho_mu}}
 \caption{ 
Observed and expected transverse energy distributions of the photon with the highest \et in the signal region in the 
\protect\subref{fig:pTpho_el}~electron and \protect\subref{fig:pTpho_mu}~muon channels of the  semileptonic analysis. 
The data are shown together with the predicted signal and backgrounds. 
Also indicated is the expected event yield for a reference model describing aQGCs with \mbox{$f_{T,0}/\Lambda^{4} =
1374\,\TeV^{-4}$} (see Section~\ref{sec:aqgc}). 
The last bin of each figure contains all overflow events.
The lower panels show the ratio of the observed number of events to the sum of expected signal and background events as well as 
the corresponding uncertainties. }
 \label{fig:semi_phoPt}
\end{figure}

Figure~\ref{fig:semi_phoPt} shows the transverse energy distributions of the photon with the highest \et in the signal region
in the electron and the muon channels. The data are shown together with the estimated background contributions and the expected
signal yield. 
The expected distribution for a reference point in the parameter space of aQGCs (see Section~\ref{sec:aqgc}) is also indicated. 
The lower panels of the two figures show the ratios of the number of observed events to the sum of expected signal and background
events.

\section{Production cross-section}
\label{sec:crosssec}
The cross-section for \wvy production is determined in fiducial regions close to the signal regions defined in
Sections~\ref{sec:fully} and~\ref{sec:semi}. While the signal region definition is based on reconstructed objects, the definition
of the fiducial region is based on particle-level MC generator information. The latter corresponds to the MC simulation
including the parton shower, hadronisation and underlying event, as opposed to the parton level, which does not account for these 
effects and solely includes the hard-scattering process of the event.

At particle level, jets are reconstructed from all stable particles (traveling at least \SI{10}{mm} before decaying) in the final
state, except for muons and neutrinos, using the \antikt algorithm with \mbox{$R = 0.4$}. 
The identification of $b$-jets at particle level is based on a matching of the jets to $b$-hadrons within a cone of size 
\mbox{$\Delta R = 0.3$} around the jet axis.
The final-state radiation of photons from leptons is accounted for by adding the four-momenta of photons that lie within a cone of
size \mbox{$\Delta R = 0.1$} around a lepton to the lepton four-momentum.
The missing transverse momentum of a particle-level event is obtained from the momenta of the neutrinos in the final state. 

The selection criteria defining the fiducial region are summarised in Table~\ref{tab:extFidReg}.
They differ from the criteria defining the signal region only for the requirements on the pseudorapidity range and the isolation
of the objects. 
Leptons are required to fulfil $|\eta| < 2.5$ and photons $|\eta| < 2.37$. 
Thus, the transition region (\mbox{$1.37 < |\eta| < 1.52$}) is included in the fiducial region and the $\eta$ requirements of the
electrons and muons are unified. 
No isolation requirements are imposed on electrons or muons. 
The photon isolation requirement is based on the isolation fraction $\epsilon_{h}^{p}$. 
The latter is defined as the ratio of the transverse energy of the closest jet that lies within a cone of size \mbox{$\Delta R = 
0.4$} around the photon to the transverse energy of the photon. 
Photons are considered isolated when $\epsilon_{h}^{p} < 0.5$.

\begin{table}[]
\centering
 \begin{tabular}{llr} \toprule \toprule
                  & \multicolumn{2}{c}{Fiducial Requirements} \\
                  & \emy                                          & $\ljjy$                       \\
    \midrule
    Leptons       & 1 electron and 1 muon                         & 1 electron or 1 muon          \\
                  & $\pt > \mbox{20\,\GeV}$                       & $\pt > \mbox{25\,\GeV}$          \\
                  & no $3^{\text{rd}}$ lepton ($\pt > \mbox{7\,\GeV}$)&  no $2^{\text{nd}}$ lepton ($\pt > \mbox{7\,\GeV}$) \\
                  & $|\eta| < 2.5$                                & $|\eta| < 2.5$                \\
                  & opposite charge  leptons                      &                               \\
                  & $\Delta R(\ell, \ell^{\prime}) > 0.1$                  &                               \\
    \midrule
    Photon        & \multicolumn{2}{c}{$\geq 1$ isolated photon }                                 \\
                  & \multicolumn{2}{c}{$\et > \mbox{15\,\GeV}$}                                      \\
                  & \multicolumn{2}{c}{isolation fraction $\epsilon_{h}^{p} < 0.5$}               \\
                  & \multicolumn{2}{c}{$|\eta| < 2.37$}                                           \\
                  & \multicolumn{2}{c}{$\Delta R(\ell, \gamma) > 0.5$}                            \\
    \midrule
    Jets          & $N_{\text{jets}} = 0$                   & $N_{\text{jets}} \geq 2$ and $N_{\text{$b$-jets}} = 0$ \\
                  & $\pt > \mbox{25\,\GeV}$                 & $\pt > \mbox{25\,\GeV}$                \\
                  & $|y| < 4.4$                             & $|\eta| < 2.5 $                     \\
                  &                                         & $|\Delta \eta_{jj}| < 1.2$   \\
                  &                                         & $\Delta R_{jj} < 3.0$        \\
                  &                                         & $\mbox{70\,\GeV} < m_{jj} < \mbox{100\,\GeV}$ \\
                  & $\Delta R(\text{jet}, \gamma) > 0.5$    & $\Delta R(\text{jet}, \gamma) > 0.5$\\
                  & $\Delta R(\text{jet}, \ell) > 0.3$      & $\Delta R(\text{jet}, \ell) > 0.3$  \\
    \midrule
    \Wboson boson & $\metRel > \mbox{15\,\GeV}$                & $\met > \mbox{30\,\GeV}$               \\
                  & $m_{e\mu}> \mbox{50\,\GeV}$                & $\mt >  \mbox{30\,\GeV}$               \\
  \bottomrule \bottomrule
  \end{tabular} 
  \caption{Definition of the fiducial regions of the fully leptonic and semileptonic \wvg analyses. 
  The objects are defined at particle level and the $\Delta R$ requirements are employed in the overlap removal. 
  The latter is implemented differently for electrons and muons. For electron--jet pairs failing the $\Delta R(\text{jet}, 
\ell)$  requirement, the jet candidate is discarded and for muon--jet pairs failing the requirement, the muon candidate is 
discarded.}
  \label{tab:extFidReg}
\end{table}

\subsection{Cross-section predictions}
\label{sec:xSecPred}

The cross-section predictions are computed at NLO in \alphas using the VBFNLO program. 
The computations are performed at parton level, while the measurement is performed at particle level. 
Therefore, the cross-section predictions are corrected to particle level by multiplying them by the parton-to-particle-level
correction factors ($C^{\text{p2p}}$). 
Each correction factor is defined as the number of signal events that satisfy the selection criteria for the fiducial region 
defined at particle level divided by the number of signal events that satisfy the selection criteria for the fiducial region 
defined at parton level. 
These factors are evaluated using the SHERPA signal simulation and amount to \mbox{$1.10 \pm 0.01$}, \mbox{$0.64 \pm 0.01$} and 
\mbox{$0.57 \pm 0.02$} for the \emy, \ejjy and \mjjy final states, respectively. 
The main difference between these corrections for the fully leptonic and the semileptonic final states arises from the 
fundamentally different requirements on the presence of jets and partons in the events.  
The difference between the electron and muon channels in the semileptonic analysis arises from different overlap removal
algorithms employed for electrons and muons; 
while jet candidates are discarded when they are close to electrons, muon candidates are discarded when they are reconstructed
close to a jet, to remove contributions from heavy-flavour quark decays. 
The uncertainties of the parton-to-particle-level correction factors include the statistical uncertainty of the SHERPA
sample and a systematic component evaluated as the difference between the corrections estimated with the SHERPA and the MadGraph
signal samples. The latter uncertainty accounts for differences in the parton shower modelling and the description of the
underlying event between the two generators.
The expected cross-section at particle level for the different final states and for the average of the electron and muon 
channels of the semileptonic analysis (\ljjy) are summarised in Table~\ref{tab:xSections}. The expected cross-sections for the
fully leptonic and semileptonic final states are of similar size despite the larger hadronic branching fraction of the \Wb and \Zb
bosons, as the selection criteria for the fiducial regions in the semileptonic analysis are more restrictive.
The uncertainty in the expected cross-section is about 5\% for all final states. 
This value accounts for the uncertainty associated with $C^{\text{p2p}}$, the numerical accuracy of the calculation, variations of
the renormalisation and factorisation scales ($\mu_{\text{R}}$ and $\mu_{\text{F}}$) by a factor of two (varied independently with
the constraint $0.5 \leq \mu_{\text{F}}/\mu_{\text{R}} \leq 2$), uncertainties due to the choice of PDF set and value of the
strong coupling constant \alphas as well as uncertainties due to the choice of isolation fraction requirement evaluated by
changing the criterion by $\pm 0.25$. 
No additional uncertainty related to the scale introduced by restricting the jet multiplicity in the fully leptonic analysis is
taken into account.
This uncertainty has been shown to be of the same order as the already included scale uncertainty by studying \Wb-boson pair production~\cite{STDM-2013-07}. Accordingly, no additional uncertainty is considered here as the experimental uncertainties are comparatively large and its inclusion would not change the results of this analysis.

\subsection{Cross-section determination}

The observed production cross-section is determined from the number of signal events in the signal region, $N_{\text{obs}}$, and
the integrated luminosity of the data set, $L_{\text{int}}$, according to  $\sigma_{\text{fid}} =
N_{\text{obs}}/(\epsilon L_{\text{int}}),$
where the correction factor, $\epsilon$, accounts for the different geometrical acceptance and selection efficiencies of the
signal region defined using reconstructed objects and the fiducial region defined at particle level.
The correction factor is evaluated using the SHERPA signal simulation and amounts to \mbox{$0.30 \pm 0.02$} for the \emy final 
state and to \mbox{$0.28 \pm 0.02$} \mbox{($0.40 \pm 0.03$)} for the electron (muon) channel of the semileptonic analysis.
The larger ranges in pseudorapidity of the leptons and photons in the fiducial region compared to the signal region contribute 
about 11\% to $\epsilon$.
The uncertainties of $\epsilon$ include the experimental uncertainties associated with the signal, a statistical component, and a 
systematic component evaluated as the difference between the corrections estimated with the SHERPA and the MadGraph signal sample 
to account for differences in the parton shower modelling and the description of the underlying event. The latter yields the 
largest contribution to the total uncertainty with the second largest contribution being the uncertainty associated with the jet 
energy scale. 

For the fully leptonic analysis, the fiducial cross-section computed using \mbox{$N^{\emy}_{\text{obs}}$} from 
Section~\ref{sec:fully} is

\begin{equation*}
 \sigma_{\text{fid}}^{\emy} = 1.5 \pm 0.9 (\text{stat.}) \pm 0.5 (\text{syst.})\,\text{fb},
\end{equation*}
where the uncertainties are symmetrised and the luminosity uncertainty is included as part of the systematic uncertainty.
The observed (expected) significance of this cross-section is determined by evaluating the $p$-value of the background-only
hypothesis at 95\% confidence level, CL, and corresponds to 1.4\,$\sigma$ sigma (1.6\,$\sigma$). The $p$-value is calculated 
using a maximum likelihood ratio as the test statistic.  
This determination of the \emy production cross-section is in agreement with the theory prediction from Table~\ref{tab:xSections}
corresponding to \SI{2.0}{fb}. The cross-section is not determined in the semileptonic final states due to its smaller
significance.

Upper limits on the production cross-sections are computed for the \emy, \ejjy and \mjjy final states and for the average 
cross-section per lepton flavour (\ljjy) in the semileptonic final states. They are determined at 95\% CL using the 
CL$_{\text{s}}$ technique~\cite{Read:2002}. 
For the combination of the semileptonic final states, the product of the likelihood functions of the \ejjy and \mjjy final
states is used as the \ljjy likelihood function in the CL$_{\text{s}}$ method.
The expected limits in the absence of a signal are computed using an Asimov data set~\cite{Cowan2010}, which provides an 
analytical approximation of the distribution of expected limits based on a $\chi^2$-distribution of the test statistics. 
The observed and expected limits are listed in Table~\ref{tab:xSections}. 
The observed limits are between 1.8 and 4.1 times larger than the SM cross-section. 
The observed upper limit on the \ljjy production cross-section is the most stringent limit reported to date. 

\begin{table}[]
\centering
\def\arraystretch{1.3}
  \begin{tabular}{lccc} 
   \toprule \toprule
                    & Observed   & Expected   & SM Prediction \\
                    & limit [fb] & limit [fb] & $\sigma_{\text{theo}}$ [fb] \\
   \midrule 
   Fully leptonic \phantom{---}$ \emy $ & 3.7 & \phantom{1}$2.1^{+0.9}_{-0.6}$ & $2.0 \pm 0.1$  \\
   Semileptonic $\left\{ \begin{tabular}{l}\ejjy\\ \mjjy \\ \ljjy \end{tabular} \right.$& 
                   \begin{tabular}{l} 10 \\ 8 \\ 6  \end{tabular}&
                   \begin{tabular}{l} \phantom{.1}$16^{+6}_{-4}$ \\ \phantom{.1}$10^{+4}_{-3}$ \\ 
\phantom{1}$8.4^{+3.4}_{-2.4}$\end{tabular}&
                   \begin{tabular}{l} $2.4 \pm 0.1$ \\ $2.2 \pm 0.1$ \\ $2.3 \pm 0.1$ \end{tabular}\\
  \bottomrule \bottomrule                                           
  \end{tabular} 
  \caption{Observed and expected cross-section upper limits at 95\% CL for the different final states using the 
CL$_{\text{s}}$ method. The expected cross-section limits are computed assuming no signal is present.
The last column shows the theory prediction for the signal cross-section ($\sigma_{\text{theo}}$) computed with the VBFNLO
program and corrected to particle level. 
The \ljjy cross-section corresponds to the average cross-section per lepton flavour in the semileptonic
analysis and all events of the \ejjy and \mjjy final states are employed for the determination of this limit.}
  \label{tab:xSections}
\end{table}

\section{Search for new physics beyond the Standard Model}
\label{sec:aqgc}
In addition to the results derived in the previous chapter, exclusion limits on the production cross-section and conficence 
intervals on aQGCs are derived in a fiducial region optimised for a search for new physics beyond the SM. 
This fiducial region differs from the fiducial region defined in Section~\ref{sec:crosssec} by an increased photon \et 
requirement. 

The aQGCs are introduced by extending the SM Lagrangian density function ($\mathcal{L}_{\text{SM}}$) with terms containing 
operators ($\mathcal{O}_{x}$) of energy-dimension eight as this is the lowest dimension that describes quartic gauge boson 
couplings without exhibiting triple gauge-boson vertices~\cite{Degrande:2013rea}.
The operators consist of different combinations of the SM fields and their coefficients are written as the
ratio of a coupling parameter ($f_{x}$) to the fourth power of the energy scale ($\Lambda$) at which the new physics 
beyond the SM would occur. Thus, the effective Lagrangian density ($\mathcal{L}_{\text{eff}}$) for \wvy production can be written 
as:

\begin{equation}\label{eq:specificAQGC}
  \mathcal{L}_{\text{eff}} = \mathcal{L}_{\text{SM}} + \sum_{j=0}^{7} \frac{f_{M,j}}{\Lambda^{4}}\mathcal{O}_{M,j}  
                                       + \sum_{j=0,1,2,5,6,7} \frac{f_{T,j}}{\Lambda^{4}}\mathcal{O}_{T,j},
\end{equation}

as there are 14 different operators that describe anomalous $WWZ\gamma$ and $WW\gamma\gamma$ couplings.
The indices $T$ and $M$ of the coupling parameter indicate two different classes of aQGC operators: 
operators containing only field strength tensors ($T$) and operators containing field strength tensors and the covariant
derivative of the Higgs field ($M$).
The SM prediction of each of the coupling parameters is zero.
The reference models in Figures~\ref{fig:fullyDistr} and~\ref{fig:semi_phoPt} depict values that are excluded by previous
analyses.

The effective field theory is not a complete model and violates unitarity at sufficiently high energy scales. 
This violation can be avoided by multiplying the coupling parameters with a dipole form factor of the form:

\begin{equation}
    \frac{1}{(1 + \hat{s}/\Lambda^2_{\text{FF}})^{2}},
\end{equation}

as described in Ref.~\cite{Eboli:2003nq}. Here, $\hat{s}$ corresponds to the squared invariant mass of the produced bosons
and $\Lambda_{\text{FF}}$ is the energy scale of the form factor. 
The latter corresponds to the energy regime above which the contributions of the anomalous couplings are largely suppressed. 
For triboson processes there is no theoretical algorithm to compute the appropriate value for $\Lambda_{\text{FF}}$ to avoid
unitarity violation. 
Therefore, the confidence intervals in this analysis are derived using three different values of $\Lambda_{\text{FF}}$:
\mbox{0.5\,\TeV},
\mbox{1\,\TeV} and infinity. The latter corresponds to the non-unitarised case, which is evaluated to allow for the comparison
with 
other analyses.

For the determination of the confidence intervals, only one coupling parameter is varied at a time and all others are set to 
zero. 
The expected number of events as a function of the varied parameter is described by a quadratic function and the predictions of
the VBFNLO program corrected to particle level are used for the determination of this function.
Confidence intervals at 95\% CL are computed using a maximum profile-likelihood ratio test statistic as done in
Ref.~\cite{STDM-2012-07}.

\begin{table}[]
  \centering 
  \def\arraystretch{1.3}
  \begin{tabular}{lcclr@{\,$\pm$\,}lr@{\,$\pm$\,}l}
  \toprule \toprule
                   & $\et^{\gamma}$ threshold [\GeV] &$N_{\text{obs}}$ & $N_{\text{bg}}$ & \multicolumn{2}{c}{$\epsilon$} &
   \multicolumn{2}{c}{$C^{\text{p2p}}$}\\ 
   \midrule
   \emy  & 120 & 0 & $0.1^{+\,0.2}_{-\,0.1}$ & 0.3 & 0.1 & 1.1 & 0.1 \\
   \ejjy & 200 & 4 & $6\phantom{1}\pm 6$             & 0.4 & 0.1 & 0.6 & 0.2 \\
   \mjjy & 200 & 3 & $4\phantom{.1}^{+\,12}_{-\,\phantom{1}4}$& 0.4 & 0.1 & 0.6 & 0.1 \\
   \bottomrule \bottomrule
  \end{tabular} 
  \caption{Numbers of observed events ($N_{\text{obs}}$) and predicted background events ($N_{\text{bg}}$) for the different final 
states with the respective photon \et threshold optimised for maximal aQGC sensitivity. 
Also given are the correction factors $\epsilon$ to correct from reconstruction level to particle level and $C^{\text{p2p}}$ to 
correct from parton level to particle level.}
  \label{tab:aQGCinputs}
\end{table}

The aQGCs would modify \wvy production at high values of $\hat{s}$ such that the sensitivity to aQGCs can be improved by raising
the threshold of the transverse energy of the photon. 
As the event count in the signal region decreases with an increasing $\et^{\gamma}$ threshold, the expected background
contribution from the other processes is extrapolated from the results obtained in Sections~\ref{sec:fully} and~\ref{sec:semi}
with \mbox{$\et^{\gamma} > \mbox{15\,\GeV}$}. 
To this end, the $\et^{\gamma}$ distribution of the total background prediction is fitted using an exponential function (the sum
of two exponential functions) in the fully leptonic (semileptonic) analysis and the total background yield is derived from the
fit. 
The optimal value of the $\et^{\gamma}$ threshold is determined by varying the threshold, computing the expected confidence 
intervals for all 14 parameters and choosing the threshold that yields the smallest expected intervals for each final state 
individually. 
This optimisation yields the best sensitivity for the requirement \mbox{$\et^{\gamma} > \mbox{120\,\GeV}$} in the fully leptonic
analysis and for \mbox{$\et^{\gamma} > \mbox{200\,\GeV}$} in both channels of the semileptonic analysis.

The number of observed events and the expected number of background events above the optimised $\et^{\gamma}$ threshold are given
in Table~\ref{tab:aQGCinputs}. The uncertainty in the background estimation includes the uncertainty in the original background
estimation and an additional uncertainty due to the extrapolation procedure, which is dominant. The latter is evaluated by varying
the fit range as well as evaluating the impact of the uncertainty of the fit parameters on the background estimation. 
Due to the higher $\et^{\gamma}$ threshold, the factors $\epsilon$ and $C^{\text{p2p}}$ are recomputed using the SM signal 
samples and are also listed in Table~\ref{tab:aQGCinputs}.
As an additional source of systematic uncertainty, $\epsilon$ and $C^\mathrm{p2p}$ are evaluated using the aQGC simulated 
samples, and their maximal deviations from the SM predictions are considered to account for their dependence on the aQGC coupling.
This uncertainty is the dominant one for $C^{\text{p2p}}$ in the fully leptonic analysis.

The upper limits on the \wvy production cross-section in the high-\et photon fiducial region are computed using the 
CL$_{\text{s}}$ formalism at 95\% CL. 
The results are given in Table~\ref{tab:xSecLimitOpti} together with limits expected in absence of \wvy production. 
In addition, the theory prediction for the SM signal cross-section computed with the VBFNLO program and corrected to
particle level is reported. The cross-section uncertainties are evaluated as described in Section~\ref{sec:xSecPred} and
range up to 22\%. 

\begin{table}[]
\def\arraystretch{1.3}
\centering
  \begin{tabular}{lcccc} 
   \toprule \toprule
                   & $\et^{\gamma}$ threshold  & Observed & Expected & SM Prediction  \\
                   &           [\GeV]           &limit [fb]&limit [fb]& $\sigma_{\text{theo}}$ [fb]\\
   \midrule 
   Fully leptonic \phantom{---}$\emy$    & 120  & 0.3 &$0.3^{+0.3}_{-0.1}$ & $0.076 \pm 0.004$ \\
   Semileptonic $\left\{ \begin{tabular}{l}\ejjy\\ \mjjy \\ \ljjy \end{tabular} \right.$ &
                    \begin{tabular}{l}  200 \\  200 \\ 200  \end{tabular}&
                    \begin{tabular}{r} 1.3 \\ 1.1 \\ 0.9  \end{tabular}&
                    \begin{tabular}{r} $1.3^{+0.5}_{-0.3}$\\$1.1^{+0.5}_{-0.3}$\\$0.9^{+0.3}_{-0.2}$\end{tabular}&
                    \begin{tabular}{l} $0.057 \pm 0.013$  \\ $0.051 \pm 0.011$   \\ $0.054 \pm 0.009$ \end{tabular}\\
  \bottomrule \bottomrule                                           
  \end{tabular} 
  \caption{Observed and expected cross-section upper limits at 95\% CL using the CL$_{\text{s}}$ method for the different final 
states with the photon \et threshold optimised for maximal aQGC sensitivity. 
The expected cross-section limits are computed assuming the absence of \wvy production. 
The last column shows the theory prediction for the SM signal cross-section computed with the VBFNLO program and corrected to 
particle level. The \ljjy cross-section corresponds to the average cross-section per lepton 
flavour in the semileptonic analysis and all events of the \ejjy and \mjjy final states are employed for the
determination of this limit.}
  \label{tab:xSecLimitOpti}
\end{table}

\begin{figure}[htbp]
 \centering
 \includegraphics[width=.7\textwidth]{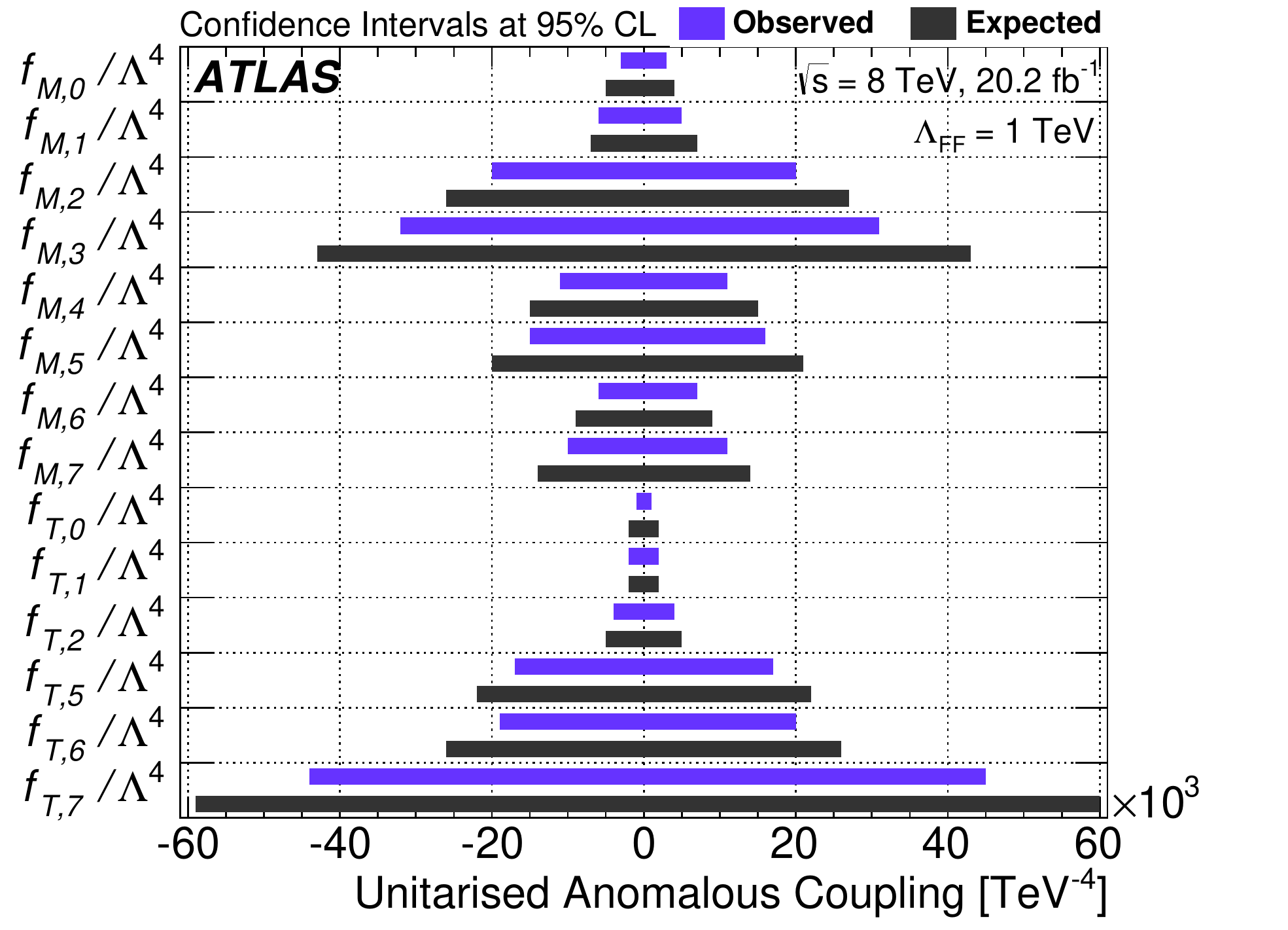}
\caption{Observed and expected confidence intervals at 95\% CL on the different anomalous quartic gauge couplings for the 
combined \wvy analysis. The couplings are unitarised using a dipole form factor with a form factor energy scale of 
\mbox{$\Lambda_{\text{FF}} = \mbox{1\,\TeV}$}.}
 \label{fig:aQGCuni1000res}
\end{figure}

For the computation of the confidence intervals, the \emy, \ejjy and \mjjy final states are combined.
The test statistic is computed from the product of the likelihood functions of the individual final states.
This combination improves the confidence intervals by up to 11\% compared to the results obtained with the \emy final state only.
The results are given in Table~\ref{tab:combinedLimitResults}.
In Figure~\ref{fig:aQGCuni1000res} the expected and observed confidence intervals using the form factor scale 
$\Lambda_{\text{FF}} = 1\,\TeV$ are shown. 
The non-unitarised couplings have also been studied by other analyses 
(e.g.~\cite{CMS-SMP-15-008, CMS-SMP-14-018, CMS-SMP-14-011, CMS-SMP-13-015, CMS-FSQ-13-008, STDM-2015-10, STDM-2015-21,
STDM-2013-05, STDM-2014-01, STDM-2015-07})
and found to be consistent with the SM prediction of zero as confirmed by this analysis.

\begin{table}[]
\centering
\def\arraystretch{1.2}
 \begin{tabular}{c|cc|cc|cc}
 \toprule \toprule
           & \multicolumn{2}{c|}{$\Lambda_{\text{FF}} = \infty$} & \multicolumn{2}{c|}{$\Lambda_{\text{FF}} = \mbox{1\,\TeV}$} & 
             \multicolumn{2}{c}{$\Lambda_{\text{FF}} = \mbox{0.5\,\TeV}$}\\
  Coupling & Observed & Expected    & Observed    & Expected    & Observed    & Expected\\
           &[$10^{3}$ \TeV$^{-4}$]&[$10^{3}$ \TeV$^{-4}$]&[$10^{4}$ \TeV$^{-4}$]&[$10^{4}$ \TeV$^{-4}$]&
            [$10^{4}$ \TeV$^{-4}$]&[$10^{4}$ \TeV$^{-4}$]\\
  \midrule
  $f_{M,0}/\Lambda^{4}$ & [$-$0.3, 0.3] &  [$-$0.4, 0.4] &  [$-$0.3, 0.3] & [$-$0.4, 0.5]& [$-$1.7, 1.8]  & [$-$2.3, 2.4]\\       
  $f_{M,1}/\Lambda^{4}$ & [$-$0.5, 0.5] &  [$-$0.8, 0.7] &  [$-$0.6, 0.5] & [$-$0.7, 0.7]& [$-$2.9, 2.6]  & [$-$4.0, 3.7]\\       
  $f_{M,2}/\Lambda^{4}$ & [$-$1.8, 1.8] &  [$-$2.4, 2.5] &  [$-$2.0, 2.0] & [$-$2.6, 2.7]& [$-$9.9, 10\phantom{.}]   & 
[\phantom{.}$-$14, 14\phantom{.}]\\     
  $f_{M,3}/\Lambda^{4}$ & [$-$3.1, 3.0] &  [$-$4.2, 4.3] &  [$-$3.2, 3.1] & [$-$4.3, 4.3]& [\phantom{.}$-$17, 16\phantom{.}]    & 
[\phantom{.}$-$23, 23\phantom{.}]\\    
  $f_{M,4}/\Lambda^{4}$ & [$-$1.1, 1.1] &  [$-$1.5, 1.6] &  [$-$1.1, 1.1] & [$-$1.5, 1.5]& [$-$5.7, 6.2]  & [$-$7.9, 8.4]\\       
  $f_{M,5}/\Lambda^{4}$ & [$-$1.7, 1.7] &  [$-$2.3, 2.3] &  [$-$1.5, 1.6] & [$-$2.0, 2.1]& [$-$8.0, 9.0]  & [\phantom{.}$-$11, 
12\phantom{.}]\\        
  $f_{M,6}/\Lambda^{4}$ & [$-$0.6, 0.6] &  [$-$0.9, 0.9] &  [$-$0.6, 0.7] & [$-$0.9, 0.9]& [$-$3.3, 3.5]  & [$-$4.7, 4.9]\\      
  $f_{M,7}/\Lambda^{4}$ & [$-$1.1, 1.1] &  [$-$1.5, 1.5] &  [$-$1.0, 1.1] & [$-$1.4, 1.4]& [$-$5.2, 5.9]  & [$-$7.5, 8.0]\\
  $f_{T,0}/\Lambda^{4}$ & [$-$0.1, 0.1] &  [$-$0.2, 0.2] &  [$-$0.1, 0.1] & [$-$0.2, 0.2]& [$-$0.9, 0.8]  & [$-$1.1, 1.1]\\
  $f_{T,1}/\Lambda^{4}$ & [$-$0.2, 0.2] &  [$-$0.2, 0.2] &  [$-$0.2, 0.2] & [$-$0.2, 0.2]& [$-$0.9, 0.9]  & [$-$1.2, 1.2]\\
  $f_{T,2}/\Lambda^{4}$ & [$-$0.4, 0.4] &  [$-$0.5, 0.5] &  [$-$0.4, 0.4] & [$-$0.5, 0.5]& [$-$1.9, 2.0]  & [$-$2.7, 2.7]\\
  $f_{T,5}/\Lambda^{4}$ & [$-$1.5, 1.6] &  [$-$2.1, 2.1] &  [$-$1.7, 1.7] & [$-$2.2, 2.2]& [$-$8.3, 8.6]  & [\phantom{.}$-$12, 
12\phantom{.}]\\ 
  $f_{T,6}/\Lambda^{4}$ & [$-$1.9, 1.9] &  [$-$2.5, 2.6] &  [$-$1.9, 2.0] & [$-$2.6, 2.6]& [\phantom{.}$-$10, 11\phantom{.}]& 
[\phantom{.}$-$14, 15\phantom{.}]\\  
  $f_{T,7}/\Lambda^{4}$ & [$-$4.3, 4.3] &  [$-$5.6, 5.8] &  [$-$4.4, 4.5] & [$-$5.9, 6.0]& [\phantom{.}$-$20, 20\phantom{.}]& 
[\phantom{.}$-$27, 28\phantom{.}]\\
  \bottomrule \bottomrule                                                        
  \end{tabular} 
  \caption{Observed and expected confidence intervals at 95\% CL on the different anomalous quartic gauge couplings for the
combined \wvy analysis for three different values of the form factor scale $\Lambda_{\text{FF}}$.}
  \label{tab:combinedLimitResults}
\end{table}

\FloatBarrier

\section{Conclusion}
\label{sec:conclusion}
The production of \wvy events is studied in \emy, \ejjy and \mjjy final states using \SI{20.2}{\per\fb} of proton--proton
collisions at a centre-of-mass energy of $\sqrt{s} = \mbox{8\,\TeV}$ recorded with the ATLAS detector at the LHC. 
The fiducial production cross-section of the \emy final state is determined with a significance of \mbox{1.4\,$\sigma$}
(\mbox{1.6\,$\sigma$} expected) and good agreement with the SM prediction at NLO in \alphas is observed.
Furthermore, upper limits on the production cross-section are derived for the \emy, \ejjy, \mjjy  and \ljjy final states in two
fiducial regions: 
one optimised for the measurement of the process and one optimised for a search for new physics beyond the SM.
No deviation from the SM predictions is observed and the results are interpreted in the framework of an effective field theory.
Confidence intervals are derived with and without unitarisation for all 14 parameters of anomalous quartic gauge couplings this 
analysis is sensitive to. 


\section*{Acknowledgements}


We thank CERN for the very successful operation of the LHC, as well as the
support staff from our institutions without whom ATLAS could not be
operated efficiently.

We acknowledge the support of ANPCyT, Argentina; YerPhI, Armenia; ARC, Australia; BMWFW and FWF, Austria; ANAS, Azerbaijan; SSTC, Belarus; CNPq and FAPESP, Brazil; NSERC, NRC and CFI, Canada; CERN; CONICYT, Chile; CAS, MOST and NSFC, China; COLCIENCIAS, Colombia; MSMT CR, MPO CR and VSC CR, Czech Republic; DNRF and DNSRC, Denmark; IN2P3-CNRS, CEA-DSM/IRFU, France; SRNSF, Georgia; BMBF, HGF, and MPG, Germany; GSRT, Greece; RGC, Hong Kong SAR, China; ISF, I-CORE and Benoziyo Center, Israel; INFN, Italy; MEXT and JSPS, Japan; CNRST, Morocco; NWO, Netherlands; RCN, Norway; MNiSW and NCN, Poland; FCT, Portugal; MNE/IFA, Romania; MES of Russia and NRC KI, Russian Federation; JINR; MESTD, Serbia; MSSR, Slovakia; ARRS and MIZ\v{S}, Slovenia; DST/NRF, South Africa; MINECO, Spain; SRC and Wallenberg Foundation, Sweden; SERI, SNSF and Cantons of Bern and Geneva, Switzerland; MOST, Taiwan; TAEK, Turkey; STFC, United Kingdom; DOE and NSF, United States of America. In addition, individual groups and members have received support from BCKDF, the Canada Council, CANARIE, CRC, Compute Canada, FQRNT, and the Ontario Innovation Trust, Canada; EPLANET, ERC, ERDF, FP7, Horizon 2020 and Marie Sk{\l}odowska-Curie Actions, European Union; Investissements d'Avenir Labex and Idex, ANR, R{\'e}gion Auvergne and Fondation Partager le Savoir, France; DFG and AvH Foundation, Germany; Herakleitos, Thales and Aristeia programmes co-financed by EU-ESF and the Greek NSRF; BSF, GIF and Minerva, Israel; BRF, Norway; CERCA Programme Generalitat de Catalunya, Generalitat Valenciana, Spain; the Royal Society and Leverhulme Trust, United Kingdom.

The crucial computing support from all WLCG partners is acknowledged gratefully, in particular from CERN, the ATLAS Tier-1 facilities at TRIUMF (Canada), NDGF (Denmark, Norway, Sweden), CC-IN2P3 (France), KIT/GridKA (Germany), INFN-CNAF (Italy), NL-T1 (Netherlands), PIC (Spain), ASGC (Taiwan), RAL (UK) and BNL (USA), the Tier-2 facilities worldwide and large non-WLCG resource providers. Major contributors of computing resources are listed in Ref.~\cite{ATL-GEN-PUB-2016-002}.

\clearpage
\printbibliography
\newpage 
\begin{flushleft}
{\Large The ATLAS Collaboration}

\bigskip

M.~Aaboud$^\textrm{\scriptsize 137d}$,
G.~Aad$^\textrm{\scriptsize 88}$,
B.~Abbott$^\textrm{\scriptsize 115}$,
O.~Abdinov$^\textrm{\scriptsize 12}$$^{,*}$,
B.~Abeloos$^\textrm{\scriptsize 119}$,
S.H.~Abidi$^\textrm{\scriptsize 161}$,
O.S.~AbouZeid$^\textrm{\scriptsize 139}$,
N.L.~Abraham$^\textrm{\scriptsize 151}$,
H.~Abramowicz$^\textrm{\scriptsize 155}$,
H.~Abreu$^\textrm{\scriptsize 154}$
% ...
\bigskip
\\
% ...
$^{12}$ Institute of Physics, Azerbaijan Academy of Sciences, Baku, Azerbaijan\\
% ...
$^{88}$ CPPM, Aix-Marseille Universit{\'e} and CNRS/IN2P3, Marseille, France\\
% ...
$^{115}$ Homer L. Dodge Department of Physics and Astronomy, University of Oklahoma, Norman OK, United States of America\\
% ...
$^{119}$ LAL, Univ. Paris-Sud, CNRS/IN2P3, Universit{\'e} Paris-Saclay, Orsay, France\\
% ...
$^{137}$ $^{(a)}$ Facult{\'e} des Sciences Ain Chock, R{\'e}seau Universitaire de Physique des Hautes Energies - Universit{\'e} Hassan II, Casablanca; $^{(b)}$ Centre National de l'Energie des Sciences Techniques Nucleaires, Rabat; $^{(c)}$ Facult{\'e} des Sciences Semlalia, Universit{\'e} Cadi Ayyad, LPHEA-Marrakech; $^{(d)}$ Facult{\'e} des Sciences, Universit{\'e} Mohamed Premier and LPTPM, Oujda; $^{(e)}$ Facult{\'e} des sciences, Universit{\'e} Mohammed V, Rabat, Morocco\\
% ...
$^{139}$ Santa Cruz Institute for Particle Physics, University of California Santa Cruz, Santa Cruz CA, United States of America\\
% ...
$^{151}$ Department of Physics and Astronomy, University of Sussex, Brighton, United Kingdom\\
% ...
$^{154}$ Department of Physics, Technion: Israel Institute of Technology, Haifa, Israel\\
$^{155}$ Raymond and Beverly Sackler School of Physics and Astronomy, Tel Aviv University, Tel Aviv, Israel\\
% ...
$^{161}$ Department of Physics, University of Toronto, Toronto ON, Canada\\
% ...
$^{a}$ Also at Department of Physics, King's College London, London, United Kingdom\\
$^{b}$ Also at Institute of Physics, Azerbaijan Academy of Sciences, Baku, Azerbaijan\\
$^{c}$ Also at Novosibirsk State University, Novosibirsk, Russia\\
$^{d}$ Also at TRIUMF, Vancouver BC, Canada\\
$^{e}$ Also at Department of Physics {\&} Astronomy, University of Louisville, Louisville, KY, United States of America\\
% ...
$^{*}$ Deceased
\end{flushleft}


\end{document}